\documentclass[11pt,a4paper]{article}
\usepackage[left=1in, right=1in, top=1in, bottom=1in]{geometry}
\usepackage{graphicx}
\usepackage{bm}
\newcommand*{\eqref}[1]{(\ref{#1})}
\usepackage[skip=8pt,font=scriptsize]{caption}
\usepackage{epstopdf}
\begin{document}
\begin{center}
{\bf Massless composite bosons formed by the coupled electron-positron pairs, and two-photon angular correlations
in the colliding beam reaction $e^{-}e^{+}\to B\gamma\gamma$ with emission of the massless boson}\\
A.I. Agafonov\\
National Research Centre "Kurchatov Institute",\\
Moscow 123182, Russia\\
Agafonov\_AIV@nrcki.ru\\
\end{center}
\begin{abstract}         
The approach in which the electron and positron are treated as ordinary, different particles, each being characterized by the complete set of the Dirac plane waves, is examined. This completely symmetric representation that is beyond the standard QED, makes it necessary to choose another solution of the Dirac equation for the free particle propagator as compared to that used currently. The Bethe-Salpeter equation with these particle propagators is studied in the ladder approximation. A new branch of the massless composite bosons formed by the coupled electron-positron pairs with the  coupling equal to the fine structure constant, has been found. It has been obtained that: 
1) the massless bosons have the normalized complex wave functions, which are transversely compressed plane waves;
2) the transverse radius of the wave functions diverges as the boson energy goes to zero that is, the composite bosons 
cannot be at rest;
3) the extension of the transverse wave function in the momentum space and the compression of the coordinate wave function are continuously occurred with increasing the boson energy.  
The reaction $e^{-}e^{+}\to B\gamma\gamma$ products of which are the massless composite boson and two photons, is investigated. The cross-section of this reaction is derived for the non-relativistic colliding beams of the spin-polarized electrons and positrons. In this case the $2\gamma$ angular correlation spectrum is characterized by a narrow peak with the full-width-at-half-maximum not exceeding 0.2 mrad. It is shown that to establish whether there is the conventional annihilation of singlet electron-positron pair with the two-photon emission or the investigated reaction, products of which are the three particles, experiments with the extremely non-relativistic colliding beams should be conducted. 
\end{abstract}
Keywords: electron-positron pairs; the Bethe-Salpeter equation; composite massless boson; two-photon angular correlation\\
PACS numbers:03.65.Pm, 11.10.St, 12.20.Ds, 12.20.-m, 13.66.Hk, 14.60.St, 14.60.Cd
\section{Introduction}	
In QED the electron and positron are currently considered as a unified particle which is described by the Dirac equation, and it is naturally assumed that the two charge states of this particle are corresponded to the positive and negative frequencies 
\cite{Fey,Akh,Ber}. The initially complete basis of the Dirac plane waves is divided into two parts: the states with positive energies are accepted as the electron states, and the states with negative energies are declared the states of the positrons which are recognized as particles traveling backwards in time \cite{Wei}. The filled states with negative energies are, as a rule, ignored, and the operator of charge conjugation, which converts the particle into antiparticle and vice versa, are introduced. Then, using the Bogoliubov-like transformation, the description of the electron-positron field can be mathematically made symmetric. However, the base of the standard model is laid initial physical asymmetry due to the introduction of the primary particles which are usually treated as the electrons, and the positrons are considered as the electronic holes in the states of the filled lower continuum \cite{Dir,Zel}. 
\par Note that in many situations, the filled electronic states with negative energies cannot be ignored. So, the filled lower continuum is important in the analysis of the electronic structure in super-heavy nuclei, for which at the certain charge of the nucleus the electron lower level 1S$_{1/2}$ merges with the bottom of the lower continuum \cite{Pop,Mig}. A similar situation arises when discussing the value of the cosmological constant. Apart from the positive contribution from the zero-point energy of boson quantized fields, another energy source is derived from the Dirac theory of the electron because the filled levels lead inevitably to negative contribution to the vacuum energy \cite{Zel}. By studying the radiation scattering by free electrons, it was concluded that radiation-induced electron quantum jumps in the intermediate states of negative energies are crucial for the scattering \cite{Tam}. 
\par Taking into account the contradictory situation presented above, in the work we examine another way in which the electron and positron are treated as independent particles, each being characterized by the complete set of the Dirac plane waves. This approach can be named as the absolutely symmetric representation of the particles. There is no reason to doubt that the complete spectrum of states for any system of interacting particles can be deduced only when the full basis of states is taken into account for each particle of the system. According to our approach the above two part division of the complete plane-wave basis leads to the following fact: neither electron states nor positron states form the complete system of the wave functions. Therefore, there is no reason to believe that all the states of the coupled electron-positron system will be obtained. 
\par The electron-positron field theory extended from the hole theory of positrons \cite{Fey}, leads to a clear picture of the annihilation process of electron-positron pairs. In this process nothing remains of the electron and positron, and the reaction products are just a few photons \cite{Fey,Ber,Dir,Ore}. So, the singlet pair of free particles with the center of mass at rest is, with the greatest probability, converted into two photons, which, due to the momentum conservation, should be emitted at the angle 180${}^{0}$ to each other. 
\par This article does not deal with high-energy electron-positron reactions, when the reaction products can be either charged leptons (electrons, muons, tau's) or hadrons \cite{Fel,Bro,Dre,Aub,Gil,Zee,Sei,Art}. The reactions of low-energy electron-positron pairs is only investigated below. In the case of the existence of these massless composite bosons, it should be a process that is, in a sense, similar to the conventional process of electron and positron annihilation, but has a fundamental difference from the latter.  This new process consists in that together with emitted photons the reaction products involve the massless boson which is formed by the strongly coupled electron-positron pair.  This annihilation-like process with emission of two photons can be as follows:
\begin{equation} 
e^{-}e^{+}\to B\gamma_{1}\gamma_{2},                                                     
\label{1} 
\end{equation}
\noindent where $B$ denotes the massless composite boson. It is essential that for this reaction  the two-photon angular correlation spectra must have finite angular widths even for the electron-positron pairs with the center of mass at rest. This is due to the fact that the three particles are the reaction products. It is of fundamental importance to find this angular width of 
the $2\gamma$ correlation spectra.
\par The article consists of two parts. The first part, presented in Section 2, is devoted to research the massless boson states formed by the coupled electron-positron system with the actual coupling equal to the fine structure constant, in the approach of the absolutely symmetric representation of the particles. It is derived that the Bethe-Salpeter equation can be reduced to the Fredholm integral equation with a non-Fredholm kernel. In general, the search for solutions of such equations is highly problematic. The problem is alleviated because the wave functions should be normalized, and the numerical procedure used to study this equation,  is given in details in details. Results obtained for the massless boson wave functions, are presented. 
\par In the case of existence of the massless composite bosons the reaction \eqref{1} should occur. The goal of the second part, presented in Section 3 of the paper, is to obtain the minimal angular width of the $2\gamma$ correlation spectrum. At first, 
the cross-section of the reaction \eqref{1} is derived for colliding  electrons and positrons with low energies. It is assumed initial pairs of free electrons and positrons with the center of mass at rest. Then the $2\gamma$ angular correlation spectrum is analyzed numerically. To establish whether there is the conventional annihilation of singlet electron-positron pair with the two-photon emission or the investigated reaction, products of which are the three particles, experiments with the extremely non-relativistic colliding beams should be conducted. The rationale for this is given in the Section 4. Here we want to note that
there is a relatively simple possibility to test the central particle-antiparticle concept of the standard model.
\par Natural units ($\hbar =c=1$) will be used throughout.
\section{The massless composite bosons}
In this Section the electron and positron are considered as independent particles. Of course, this approach does not correspond to the generally accepted particle-antiparticle concept in QED. However, from the theoretical point it seems important to investigate this way as well and understand what can be expected from this consideration. 
\subsection{The free fermion propagator}
At present in QED the free fermion propagator for the Dirac equation is used in the form \cite{Fey,Akh,Ber}:
\begin{equation}  
K_{+}(2,1)=\sum_{\bf p}\psi_{p}(2)\bar{\psi }_{p}(1)\theta (t_{2} -t_{1} )-\sum_{\bf p}\psi_{-p} (2)\bar{\psi }_{-p}  
\theta (t_{1}-t_{2}).                          
\label{2}
\end{equation} 
\noindent Here $\psi_{\pm p}$ is the Dirac plane wave representing the state of the free particle with energy 
$\pm \varepsilon_{\bf p} $, respectively, and  $\bar{\psi }_{p} =\psi_{p}^{*} \beta$ denotes the Dirac conjugation. 
In Eq. \eqref{2} the contribution to $K_{+} (2,1)$ at $t_{2}>t_{1}$  is due to the electron terms, and at $t_{2}<t_{1} $ - the positron terms.
\par The Bethe-Salpeter equation \cite{Sal,Gel} with the propagator \eqref{2} has been studied in many works, as a rule, in the ladder approximation. The results obtained can be summarized as follows. Firstly, the weakly bound states of positronium have been found with certain relativistic corrections. Secondly, after the work \cite{Gol}, considerable interest is attached to the problem of strongly coupled states for fermion-antifermion systems. In the most commonly used approach to the problem the Bethe-Salpeter equation is regarded as eigenvalues task for the coupling constant \cite{Nak,Set,Nis,Sut,Luc,Lad,Baw,Saz}. That is, 
an eigenvalue is considered as the necessary strength of the attractive potential to make a massless bound state.
\par Thirdly, the Feynman theory leads to a clearer picture of the annihilation of electron-positron pairs \cite{Fey,Wei}. In principal this process can be analyzed by using the Bethe-Salpeter equation with account for the ``annihilation'' interaction which can be treated as a perturbation \cite{Gre}. Therefore, any theory of the electron-positron field must include a process that at least looks the annihilation-like process.
\par The Dirac equation, as well as any differential equation, has several solutions for Green's function \cite{Iva}. For example, the retarded Green function of the following form was also discussed in \cite{Fey}:
\begin{equation} 
K_{-}(2,1)=\sum_{\bf p}\left(\psi_{p} (2)\bar{\psi}_{p} (1)+\psi_{-p} (2)\bar{\psi}_{-p} (1)\right) \theta (t_{2}-t_{1}).
\label{3} 
\end{equation} 
\par Comparing \eqref{2} and \eqref{3}, one can conclude that unlike the propagator \eqref{3}, the negative energy states are actually excluded from consideration in \eqref{2}. Treating \eqref{3} as the electron propagator, we can see that this function is taken into account all the spectrum of the Dirac plane waves. There is no doubt that the positron can be described by the Dirac equation as well. Then the only opportunity to use a free propagator similar to \eqref{3} for the positron is to assume that the electron and positron are independent particles. For the positron propagator we would have:
\begin{equation}
K_{+}(4,3)=\sum_{\bf p}\left(\varphi_{p}(4)\bar{\varphi}_{p}(3)+\varphi_{-p}(4)\bar{\varphi}_{-p}(3)\right)\theta (t_{4}-t_{3}).
\label{4} 
\end{equation} 
In Eqs. \eqref{3} and \eqref{4} $\psi_{\pm p}$ and $\varphi_{\pm p}$ are the Dirac plane waves for the free electrons and positrons, respectively. 
\par In this approach, in the vacuum state the lower continua for each of these particles are completely filled and the upper continua are not occupied. Then this vacuum state is charge neutral.
\par The interaction between electron and positron is attractive. In the ladder approximation the retarded interaction function 
can be written as \cite{Fen}:
\begin{equation} 
G^{(1)}(3,4;5,6)=-e^{2}(1-{\bm \alpha}_{-}{\bm \alpha}_{+})\delta_{+}(s_{56}^{2})\delta (3,5)\delta (4,6).
\label{5}
\end{equation} 
\noindent Here ${\bm \alpha}_{\pm }=\left(\begin{array}{cc} {0} & {{\bm \sigma}_{\pm } } \\ {{\bm \sigma}_{\pm } } & {0} \end{array}\right)$are the velocity operators for the electron (-) and positron (+),  and ${\bm \sigma}_{\pm }$ are the Pauli matrices for the electron and positron, and $s_{56} $ is the invariant distance between the particles. 
\par The propagators \eqref{3} and \eqref{4} are taken into account the complete orthonormal system of the wave functions for each of the particles. However, according to \eqref{5}, the interaction between them is attractive only if both these particles are in the states with the positive (negative) energies. If one of them is in the states of the upper continuum, and the other - in the states of the lower continuum, the interaction between the particles changes sign, and becomes repulsive. Therefore propagators \eqref{3} and \eqref{4} are not suitable. 
\par Taking into account for \eqref{5}, in the approach of the absolutely symmetric representation of the particles the electron retarded Green function should be written as:
\begin{equation} \label{6} 
K_{0-}(2,1)=\sum_{\bf p}\left(\psi_{p}(2)\psi_{p}^{+} (1)+\psi_{-p}(2)\psi_{-p}^{+}(1)\right)\theta (t_{2}-t_{1}),
\end{equation} 
\noindent and, similarly, for the positron propagator 
\begin{equation}\label{7}  
K_{0+}(4,3)=\sum_{\bf p}\left(\varphi_{p} (4)\varphi_{p}^{+} (3)+\varphi_{-p}(4)\varphi_{-p}^{+}(3)\right)\theta (t_{4}-t_{3}).
\end{equation} 
Here $\psi _{\pm p}^{+} $ and $\varphi _{\pm p}^{+} $ are the Hermitian conjugate matrices with respect to $\psi _{\pm p}^{} $ and $\varphi _{\pm p}^{} $, respectively. The letter are given by: $\psi _{p} ,\varphi _{p} =u_{p,\pm } e^{-ipx} $and $\psi _{-p} ,\varphi _{-p} =u_{-p,\pm } e^{ipx} $, where
\begin{equation}  
u_{p,\pm } =\frac{1}{\sqrt{2\varepsilon _{p} } } \left(\begin{array}{c} {\sqrt{\varepsilon _{p} +mc^{2} } w_{\pm } } \\ {\sqrt{\varepsilon _{p} -mc^{2} } ({\bf n}{\bm \sigma}_{\pm } )w_{\pm} } \end{array}\right)u_{-p,\pm} =\frac{1}{\sqrt{2\varepsilon _{p} } } \left(\begin{array}{c} {\sqrt{\varepsilon _{p} -mc^{2} } ({\bf n}{\bm \sigma}_{\pm } )w_{\pm }^{'} } \\ {\sqrt{\varepsilon _{p} +mc^{2} } w_{\pm }^{'} } \end{array}\right),               
\label{8}
\end{equation} 
and $\bf n$ is the unit vector, ${\bf n}={\bf p}/p$.
\par Using \eqref{6}-\eqref{8}, in the ladder approximation the bound-state Bethe-Salpeter equation for the electron-positron system is:
\begin{equation}\label{9}  
\psi (1,2)=-i\int \int \int \int d\tau _{3}d\tau _{4}d\tau _{5}d\tau _{6}K_{0-}(1,3)K_{0+}(2,4)G^{(1)}(3,4;5,6)\psi (5,6),
\end{equation} 
where $d\tau _{i}=d{\bf r}_{i}dt_{i}$.
\par Below we do not interest the positronium states. Note only that in the non-relativistic limit in which one neglects the interaction retardation and the interaction through the vector potential, and assumes that the characteristic velocity of particles in the bound pair is much smaller than the speed of light, Eq. \eqref{9} with the propagators \eqref{6}-\eqref{7} is reduced to the Schrodinger equation for the Ps states.
\subsection{The boson wave function}
We are searching for a solution $\psi (1,2)$ of Eq. \eqref{9} in the form of a stationary wave with the phase velocity equal to the speed of light. Let ${\bf p}+{\bf q}={\bf g}$ (where $\bf p$ and $\bf q$ are the momentum of the electron and positron), and the momentum of the pair, $\bf g$, is directed along the {\it z}-axis. It is the strongly coupled state with the energy $E=g$. Due to the symmetry of the problem, we have to put $z_{1}=z_{2}=z$ and $t_{1}=t_{2}=t$ for this massless boson state that allows us to introduce the two-dimensional relative vector between the particles, ${\bm \rho}={\bm \rho}_{1}-{\bm \rho}_{2} $. Then the wave function is: 
\begin{equation}\label{10}  
\psi (1,2)=\varphi({\bm \rho},{\bf g})\exp (ig(z-t)).                                                 
\end{equation} 
\par One can imagine \eqref{10} as transversely compressed plane wave. The wave cross-section is determined by the wave function of the transverse motion of the coupled pair, $\varphi({\bm \rho},{\bf g})$. The latter satisfies the normalization:
\begin{displaymath}
\int |\varphi({\bm \rho},{\bf g})|^{2}d{\bm \rho} =1. 
\end{displaymath}
\par The particle distribution is stationary, and depends only on $\bm \rho$.
\par The function $\delta _{+}(s_{56}^{2})$ contained in \eqref{5}, was given in \cite{Fen}. In our case it can be written as:  
\begin{equation}  
\delta_{+}(t_{56}^{2}-\rho _{56}^{2} )=\frac{1}{4\pi \rho _{56} }\int_{-\infty}^{+\infty}\left(e^{-i\omega (t_{5}-t_{6} )} 
+e^{-i\omega (t_{6}-t_{5} )}\right)e^{i|\omega |\rho _{56} }d\omega,                     
\label{11}
\end{equation} 
where $\rho _{56} =|{\bm \rho}_{5} -{\bm \rho}_{6}|$. It was taken into account that since the phase velocity of the wave 
\eqref{10} is equal to the velocity of light, for the stationary distribution of the particle density the interaction between 
the electron and positron can only occur in the same layers ($z_{5} =z_{6}$), which are perpendicular to the wave vector $\bf g$. Considering that $\alpha_{-z}\alpha_{+z} =1$ (here $\alpha_{-z}$ and $\alpha_{+z}$ are the \textit{z}- component of the velocity operators for the electron and positron, respectively), the factor of $(1-{\bm \alpha}_{-}{\bm \alpha}_{+})$ in \eqref{5} should be replaced by $-{\bm \alpha}_{-\rho }{\bm \alpha}_{+\rho }$. 
\par The latter means that in this bound state the electron and positron do not interact through the Coulomb potential and their retarded interaction occurs through the vector potential which is due to the particles transverse motion defined by the function
$\varphi({\bm \rho},{\bf g})$.
\par As a result, for the massless composite boson state \eqref{10} Eq. \eqref{9} is reduced to:
\begin{displaymath}  
\varphi ({\bm \rho}_{12} ,{\bf g})e^{igz-igt} =-ie^{2} \int_{-\infty }^{z}dz_{3} \int _{-\infty }^{z}dz_{4} 
\int d{\bm \rho}_{3} \int d{\bm \rho}_{4} \int _{-\infty }^{t}dt_{3} \int _{-\infty }^{t}dt_{4}
\end{displaymath}  
\begin{displaymath}
\sum _{{\bf pq}}\frac{ e^{ \left(i{\bf p}({\bf r}_{1} -{\bf r}_{3} )+i{\bf q}({\bf r}_{2} -{\bf r}_{4} )\right)} }
{4\varepsilon _{p} \varepsilon _{q}} 
\int _{-\infty }^{+\infty }e^{i|\omega |\rho _{34} } d\omega \left\{\Lambda _{-}^{+} ({\bf p})e^{-i\varepsilon _{p} (t-t_{3} )} +
\Lambda _{-}^{-} ({\bf p})e^{i\varepsilon _{p} (t-t_{3} )} \right\}
\end{displaymath}  
\begin{displaymath}
\left\{\Lambda_{+}^{+} ({\bf q})e^{-i\varepsilon_{q}(t-t_{4})} +\Lambda_{+}^{-} ({\bf q})e^{i\varepsilon_{\bf q} (t-t_{4} )} \right\} 
\left\{e^{-i\omega (t_{3} -t_{4} )} +e^{-i\omega (t_{4} -t_{3} )} \right\}
\end{displaymath}  
\begin{equation}
e^{i\frac{g}{2} (z_{3} +z_{4})-i\frac{g}{2} (t_{3} +t_{4} )} \frac{1}{4\pi \rho_{34} } 
\left({\bm \alpha}_{-\rho } {\bm \alpha}_{+\rho} \right)\varphi ({\bm \rho}_{34} ,{\bf g)}  
\label{12}
\end{equation} 
Here $\Lambda_{-}^{+}({\bf p})=\varepsilon_{{\bf p}}+m\beta_{-} +{\bm \alpha}_{-}{\bf p}_{\rho} +p_{z}$ and 
$\Lambda_{-}^{-}({\bf p})=\varepsilon_{{\bf p}} -m\beta_{-} -{\bm \alpha}_{-} {\bf p}_{\rho }-p_{z}$ are the electron operators, 
$\Lambda_{+}^{+}({\bf q})=\varepsilon_{{\bf q}} +m\beta_{+} +{\bm \alpha}_{+} {\bf q}_{\rho }+q_{z}$ and 
$\Lambda_{+}^{-}({\bf q})=\varepsilon_{{\bf q}} -m\beta_{+} -{\bm \alpha}_{+} {\bf q}_{\rho }-q_{z}$ are the positron ones.
\par At first, analyzing only the \textit{z-} dependent functions in \eqref{12}, we integrate over $z_{3}$ and $z_{4}$:
\begin{displaymath}
e^{igz} =\sum_{p_{z} q_{z} }\int _{-\infty }^{z\to \infty }dz_{3} \int_{-\infty }^{z\to \infty }dz_{4} 
e^{i\frac{g}{2} (z_{3} +z_{4} )+ip_{z} (z-z_{3} )+iq_{z} (z-z_{4})}  
\end{displaymath}  
\begin{displaymath}
=(2\pi )^{2} \sum _{p_{z} q_{z} } e^{ i(p_{z} +q_{z} )z}\delta (p_{z} -\frac{g}{2} )\delta (q_{z} -\frac{g}{2} )=e^{igz} 
 \left|_{q_{z} =p_{z} =\frac{g}{2}}\right.  
\end{displaymath}  
\par Since the function $\varphi$ depends only on ${\bm \rho}_{34}$, in \eqref{12} we replace the integration variables: 
$\int d{\bm \rho}_{3}\int d{\bm \rho}_{4}=\int d{\bm \rho}_{34} \int d({\bm \rho}_{3} +{\bm \rho}_{4})/2$. 
Thereafter the integral over $({\bm \rho}_{3} +{\bm \rho}_{4} )/2$ on the right side of \eqref{12} is easily calculated,  
and gives $(2\pi )^{2} \delta ({\bf p}_{\rho } +{\bf q}_{\rho })$. Consequently Eq. \eqref{12} takes the form:
\begin{displaymath}
\varphi ({\bm \rho} _{12} )e^{-igt} =-i\frac{e^{2} }{4\pi } \int \frac{d{\bm \rho} _{34} }{\rho_{34} } 
\int _{-\infty }^{t}dt_{3}  \int _{-\infty }^{t}dt_{4} \sum _{{\bf p}_{\rho } }\frac{e^{i{\bf p}_{\rho } ({\bf r}_{12} -{\bf r}_{34}) }}
{4\varepsilon _{p}^{2} }  \int _{-\infty }^{+\infty }e^{i|\omega |\rho _{34} }  d\omega 
\end{displaymath}  
\begin{displaymath}
\left\{\Lambda _{-}^{+} ({\bf p})e^{-i\varepsilon_{p} (t-t_{3} )} +\Lambda _{-}^{-} ({\bf p})e^{i\varepsilon _{p} (t-t_{3} )} \right\}
\left\{\Lambda _{+}^{+} ({\bf q})e^{-i\varepsilon_{q} (t-t_{4} )} +\Lambda _{+}^{-} ({\bf q})e^{i\varepsilon _{q} (t-t_{4} )} \right\}
\end{displaymath}  
\begin{equation}  
\left\{e^{-i\omega (t_{3} -t_{4} )} +e^{-i\omega (t_{4} -t_{3} )} \right\}e^{-i\frac{g}{2} (t_{3} +t_{4} )} 
\left({\bm \alpha}_{-\rho } {\bm \alpha}_{+\rho } \right)\varphi ({\bm \rho}_{34},{\bf g}),
\label{13}
\end{equation} 
where ${\bf q}_{\rho }=-{\bf p}_{\rho }$, $q_{z}=p_{z} =\frac{g}{2}$ and $\varepsilon_{p}=\varepsilon _{q}$.
\par Now integrating over $t_{3}$ and $t_{4}$, Eq. \eqref{13} is rewritten as:
\begin{equation}  
\varphi ({\bm \rho}_{12},{\bf g})=-i\frac{e^{2} }{4\pi} \int \frac{d {\bm \rho}_{34} }{\rho_{34}} \sum_{{\bf p}_{\rho}}
\frac{e^{ i{\bf p}_{\rho}({\bf \rho}_{12} -{\bf \rho}_{34})}}{4\varepsilon_{p}^{2}}  
\int_{-\infty }^{\infty }e^{i|\omega |\rho _{34} }d\omega ({\rm I}_{1}+{\rm I}_{2}+{\rm I}_{3})
\left({\bm \alpha}_{-\rho}{\bm \alpha}_{+\rho }\right)\varphi({\bm \rho}_{34},{\bf g}),         
\label{14}
\end{equation} 
\noindent where the functions ${\rm I}_{1,2,3}(\omega)$ are given by:
\[{\rm I}_{1} =2\frac{\Lambda_{-}^{+}({\bf p})\Lambda_{+}^{+}({\bf q})}{\omega^{2}-(\varepsilon_{p}-\frac{g}{2})^{2}+i\delta}, 
 {\rm I} _{2} =2\frac{\Lambda_{-}^{-}({\bf p})\Lambda_{+}^{-}({\bf q})}{\omega^{2}-(\varepsilon_{q}+\frac{g}{2})^{2}-i\delta}\] 
and
\[{\rm I}_{3} =\frac{\Lambda_{-}^{-}({\bf p})\Lambda_{+}^{+}({\bf q})+\Lambda_{-}^{+}({\bf p})\Lambda_{+}^{-}({\bf q})}{-g} 
\left(\frac{2\varepsilon_{p}-g}{\omega^{2}-(\varepsilon_{p}-\frac{g}{2})^{2}+i\delta}+\frac{-2\varepsilon_{p}-g}{\omega^{2}
 -(\varepsilon_{p}+\frac{g}{2})^{2}-i\delta}\right). \] 
\noindent Here $\delta \to 0^{+} $ is defined the rule for bypassing simple poles. 
\par All the three integrals over $\omega $ on the right side of \eqref{14} are easily calculated. Finally we obtain:
\begin{equation}  
T_{1} =\int_{-\infty }^{\infty}{\rm I}_{1}(\omega )e^{i|\omega |\rho_{34}}d\omega =
8i\frac{\Lambda_{-}^{+}({\bf p})\Lambda_{+}^{+}({\bf q})}{2\varepsilon_{p}-g} \left(\cos(x)si(x)-\sin(x)ci(x)\right),
\label{15}
\end{equation} 
where $x=|\varepsilon_{p}-\frac{g}{2}|\rho_{34}$,
\begin{equation} 
T_{2} =\int_{-\infty }^{\infty}{\rm I}_{2}(\omega)e^{i|\omega |\rho_{34}}d\omega =
8i\frac{\Lambda_{-}^{-}({\bf p})\Lambda_{+}^{-}({\bf q})}{2\varepsilon_{p}+g}\left\{\pi e^{iy}+\cos(y)si(y)-\sin(y)ci(y)\right\} 
\label{16} 
\end{equation} 
with $y=(\varepsilon_{p}+\frac{g}{2} )\rho_{34}$ and
\begin{displaymath}
T_{3} =\int_{-\infty }^{\infty}{\rm I}_{3}(\omega )e^{i|\omega |\rho_{34}}d\omega =
4i\frac{\Lambda_{-}^{-}({\bf p})\Lambda_{+}^{+}({\bf q})+\Lambda_{-}^{+}({\bf p})\Lambda_{+}^{-}({\bf q})}{g}  
\end{displaymath}
\begin{equation}  
\left\{\pi e^{iy} -\cos(x)si(x)+\sin(x)ci(x)+\cos(y)si(y)-\sin(y)ci(y)\right\}.                          
\label{17}
\end{equation} 
Here $si(x)$ and $ci(x)$ are the integral sine and cosine. 
\par Now we are concerned with the bispinors of the function $\varphi$. It is apparent from \eqref{14} that the action of the following operators $\alpha_{-x} \alpha_{+x} +\alpha_{-y} \alpha_{+y} $,  $\Lambda_{-}^{+} ({\bf p})\Lambda_{+}^{+} ({\bf q})$, 
$\Lambda_{-}^{-} ({\bf p})\Lambda_{+}^{-} ({\bf q})$  and $\Lambda_{-}^{-} ({\bf p})\Lambda_{+}^{+} (q)+
\Lambda _{-}^{+} (p)\Lambda _{+}^{-} (q)$ on these bispinors must be reduced only to the multiplication of these spin functions on some scalars. For the boson state each particle of the bound pair can be characterized by the projection of the particle spin on the wave vector $g$ or, in other words, the particle helicity. There are eight bispinor functions $\eta _{i} (i=1,...,8)$ for which the helicities of both the electron and positron are simultaneously either positive 
\[\eta _{i=1,...,4} =\left(\begin{array}{c} {1} \\ {0} \\ {0} \\ {0} \end{array}\right)\left(\begin{array}{c} {1} \\ {0} \\ {0} \\ {0} \end{array}\right),\left(\begin{array}{c} {1} \\ {0} \\ {0} \\ {0} \end{array}\right)\left(\begin{array}{c} {0} \\ {0} \\ {1} \\ {0} \end{array}\right),\left(\begin{array}{c} {0} \\ {0} \\ {1} \\ {0} \end{array}\right)\left(\begin{array}{c} {1} \\ {0} \\ {0} \\ {0} \end{array}\right),\left(\begin{array}{c} {0} \\ {0} \\ {1} \\ {0} \end{array}\right)\left(\begin{array}{c} {0} \\ {0} \\ {1} \\ {0} \end{array}\right),\] 
or negative
\[\eta _{i=5,...,8} =\left(\begin{array}{c} {0} \\ {1} \\ {0} \\ {0} \end{array}\right)\left(\begin{array}{c} {0} \\ {1} \\ {0} \\ {0} \end{array}\right),\left(\begin{array}{c} {0} \\ {1} \\ {0} \\ {0} \end{array}\right)\left(\begin{array}{c} {0} \\ {0} \\ {0} \\ {1} \end{array}\right),\left(\begin{array}{c} {0} \\ {0} \\ {0} \\ {1} \end{array}\right)\left(\begin{array}{c} {0} \\ {1} \\ {0} \\ {0} \end{array}\right),\left(\begin{array}{c} {0} \\ {0} \\ {0} \\ {1} \end{array}\right)\left(\begin{array}{c} {0} \\ {0} \\ {0} \\ {1} \end{array}\right).\] 
Because ${\bm\alpha}_{-\rho } {\bm \alpha}_{+\rho } =\alpha _{-x} \alpha _{+x} +\alpha _{-y} \alpha _{+y} $, one can convince that
\[{\bm \alpha}_{-\rho } {\bm \alpha}_{+\rho } \eta _{i} =0.\] 
\par It implies that the interaction function $G^{(1)}(3,4;5,6)\propto \alpha_{-\rho }\alpha_{+\rho }$ vanishes, and the massless boson state cannot be formed for these states.
\par Analyzing the remaining possibilities, the required functions can be written as:
\begin{displaymath}  
\varphi_{1} ({\bm \rho} )=\chi _{1} ({\bm \rho} )\left[\left(\begin{array}{c} {1} \\ {0} \\ {0} \\ {0} \end{array}\right)_{-} \left(\begin{array}{c} {0} \\ {0} \\ {0} \\ {1} \end{array}\right)_{+} +\left(\begin{array}{c} {0} \\ {0} \\ {0} \\ {1} \end{array}\right)_{-} 
\left(\begin{array}{c} {1} \\ {0} \\ {0} \\ {0} \end{array}\right)_{+} \right],
\end{displaymath}  
\begin{equation}  
\varphi_{1} ({\bm \rho})=\chi _{2} ({\bm \rho})\left[\left(\begin{array}{c} {0} \\ {1} \\ {0} \\ {0} \end{array}\right)_{-} \left(\begin{array}{c} {0} \\ {0} \\ {1} \\ {0} \end{array}\right)_{+} +\left(\begin{array}{c} {0} \\ {0} \\ {1} \\ {0} \end{array}\right)_{-} \left(\begin{array}{c} {0} \\ {1} \\ {0} \\ {0} \end{array}\right)_{+} \right].                
\label{18}
\end{equation} 
Here $\chi_{1,2}({\bm \rho},{\bf g})$ are the coordinate wave functions of the transverse motion of the strongly coupled 
electron-positron pair. 
\par In the states \eqref{18} the helicities of the electron and positron are opposite. For the functions \eqref{18} we have:
\[{\bm \alpha}_{-\rho } {\bm \alpha}_{+\rho } \varphi _{1,2} =2\varphi _{1,2} , \] 
\[\Lambda _{-}^{+} ({\bf p})\Lambda _{+}^{+} ({\bf q})\varphi _{1,2} =g(\varepsilon _{p} +\frac{g}{2} )\varphi _{1,2} ,\] 
\[\Lambda _{-}^{-} ({\bf p})\Lambda _{+}^{-} ({\bf q})\varphi _{1,2} =g(-\varepsilon _{p} +\frac{g}{2} )\varphi _{1,2} ,\] 
\[(\Lambda _{-}^{-} ({\bf p})\Lambda _{+}^{+} ({\bf q})+\Lambda _{-}^{+} ({\bf p})\Lambda _{+}^{-} ({\bf q}))\varphi _{1,2} 
=\left(4\varepsilon _{p}^{2} -g^{2} \right)\varphi _{1,2} .\] 
\par As a result, Eq. \eqref{14} is transformed to the following integral equation for the coordinate function $\chi_{1,2}$ 
(since $\chi_{1}=\chi_{2}$ the lower index 1,2 can be omitted):
\begin{equation}  
\chi ({\bm \rho}_{12})=\frac{e^{2}}{(2\pi)^{3}}\int \frac{d{\bf \rho}_{34}}{\rho_{34}}\int \frac{d{\bf p}_{\rho}}
{\varepsilon_{p}^{2}} \exp\left(i{\bf p}_{\rho}({\bm \rho}_{12}-{\bm \rho}_{34})\right)\left(T_{1}+T_{2}+T_{3}\right)
\chi({\bm \rho}_{34}),                 
\label{19}
\end{equation} 
where
\begin{equation}  
T_{1} =g\frac{2\varepsilon_{p}+g}{2\varepsilon_{p}-g} \left(\cos(x)si(x)-\sin(x)ci(x)\right),
\label{20}
\end{equation} 
\begin{equation}  
T_{2} =g\frac{g-2\varepsilon_{p}}{g+2\varepsilon_{p}} \left\{\pi e^{iy} +\cos(y)si(y)-\sin(y)ci(y)\right\} 
\label{21}
\end{equation} 
and
\begin{equation}  
T_{3} =\frac{4\varepsilon_{p}^{2}-g^{2}}{g} \left\{\pi e^{iy} -\cos(x)si(x)+\sin(x)ci(x)+\cos(y)si(y)-\sin(y)ci(y)\right\}.
\label{22}
\end{equation} 
\par Eq. \eqref{19} with the notations \eqref{20}-\eqref{22} is a rather complicated integral equation for the transverse wave function 
$\chi({\bm \rho}_{12},{\bf g})$. It is important to show that this equation has solutions for the normalized eigenfunctions. 
Below it will demonstrate for the case of the \textit{S}-state of the bound pair and small momenta of the boson, $g<<m$.
\par For the \textit{S}-state the angular momentum of the relative motion of the bound pair is zero, and the transverse wave function depends only on the modulus of the relative vector, that is $\chi(\rho_{12},g)$. In such an event after integration over the azimuthal angle of the vector ${\bm \rho}_{34}$ and integration over the azimuthal angle of the vector ${\bf p}_{\rho }$  
Eq. \eqref{19} is reduced to:
\begin{equation}  
\chi(\rho_{12})=\frac{e^{2} }{2\pi} \int_{0}^{\infty}d \rho_{34} \int_{0}^{\infty}p_{\rho} dp_{\rho}  J_{0}(p_{\rho} 
\rho_{12})J_{0}(p_{\rho} \rho_{34})\frac{\sum_{i=1}^{3}T_{i}(q,\rho_{34})}{\varepsilon_{q}^{2}} \chi(\rho_{34}).
\label{23}
\end{equation} 
Here $J_{0}$ is the Bessel function of the first kind. In the case $g<<m$ from \eqref{19}-\eqref{21} we find that 
$T_{3} >>T_{1},T_{2} $ and 
\begin{equation}  
T_{3} \cong 4\pi \frac{\varepsilon_{q}^{2}}{g}\exp (i\varepsilon_{q}\rho_{34}).
\label{24}
\end{equation} 
Substituting \eqref{24} to Eq. \eqref{23}, the latter equation is reduced to the homogeneous Fredholm integral equation of the second kind:       
\begin{equation} 
\chi(\rho_{12})=2\frac{e^{2}}{g} \int_{0}^{\infty}d \rho_{34} R(\rho_{12},\rho_{34})\chi(\rho_{34}) 
\label{25} 
\end{equation} 
with the kernel
\begin{equation} 
R(\rho_{12},\rho_{34})=\int_{0}^{\infty}p_{\rho} dp_{\rho} J_{0}(p_{\rho}\rho_{12})J_{0}(p_{\rho}\rho_{34})
\exp (i\varepsilon_{p} \rho_{34}).                       
\label{26} 
\end{equation} 
\par One can convince that the kernel \eqref{26} is non-Fredholm one. According to the asymptotic property of the Bessel function 
$J_{0}$, the integral in the right-hand side of \eqref{26} can only be defined as the principal value integral. That is, the equation \eqref{25} with the kernel \eqref{26} is still difficult to study. 
\par In the momentum space, the equation for the transverse wave function can be reduced to a simpler form convenient for its numerical solving. For this, in \eqref{19} we use the Fourier transform of the $\chi$-functions,  
$\chi ({\bf q})=\int \chi({\bm \rho}_{12})\exp(-i{\bf q}{\bm \rho}_{12})d{\bm \rho}_{12}$. Then, for the wave function in the momentum space we obtain:
\begin{equation} 
\chi({\bf q})=\frac{e^{2}}{(2\pi)^{3}} \int \frac{d{\bm \rho}_{34}}{\rho_{34}} \frac{\sum_{i=1}^{3}T_{i} (q,\rho_{34})}
{\varepsilon_{q}^{2}} \int d{\bf f} e^{i({\bf f}-{\bf q}){\bm \rho}_{34}} \chi ({\bf f}).                          
\label{27} 
\end{equation} 
For the \textit{S}-state of the bound pair Eq. \eqref{27} can be written as:
\begin{equation}  
\chi(q)=\frac{e^{2}}{2\pi \varepsilon_{q}^{2}} \int d\rho_{34} \sum _{i=1}^{3}T_{i} (q,\rho_{34}) J_{0}(q\rho_{34})
\int f J_{0} (f\rho_{34}) \chi(f)df.                     
\label{28}
\end{equation} 
In the case $g<<m$ the function $\sum _{i=1}^{3}T_{i} (q,\rho_{34})$ can be replaced by the expression \eqref{24}. Then Eq. 
\eqref{28} is reduced to the homogeneous Fredholm integral equation of the second kind:
\begin{equation} 
\chi(q)=\frac{e^{2}}{g} \int_{0}^{\infty}Q(q,f)\chi(f)df  
\label{29} 
\end{equation} 
with the kernel                                        
\begin{equation} 
Q(q,f)=2f\int_{0}^{\infty}J_{0}(q\rho_{34})J_{0}(f\rho_{34})\exp (i\varepsilon_{q}\rho_{34})d\rho_{34}.
\label{30} 
\end{equation} 
\par Now the integral in the right-hand side of \eqref{30} that has a direct relationship to the discontinuous Weber-Schafheitlin integral, is absolutely convergent integral which, however, is expressed through a discontinuous function, as it will see slightly below. 
\par It is convenient to use the dimensionless variables: 
$x\to \rho_{34} /{\mathchar'26\mkern-10mu\lambda}_{e}$, $q\to q{\mathchar'26\mkern-10mu\lambda}_{e}$, 
$f\to f{\mathchar'26\mkern-10mu\lambda}_{e}$ and $g\to gm$. Here ${\mathchar'26\mkern-10mu\lambda}_{e}$ is the Compton wave length of the electron. Then Eqs. \eqref{29}-\eqref{30} are rewritten as:
\begin{equation}  
\chi(q)=\frac{\alpha}{g} \int_{0}^{\infty}Q(q,f)\chi(f)df  
\label{31}
\end{equation} 
and
\begin{equation} 
Q(q,f)=2f\int_{0}^{\infty} J_{0}(qx)J_{0}(fx)\exp (ix\sqrt{1+q^{2}})dx .                                    
\label{32} 
\end{equation} 
Here $\alpha$ is the fine structure constant.  
\par The integral on the right side of \eqref{32} was previously calculated \cite{Pru}. Thus the complex kernel $Q(q,f)=ReQ+iImQ$ is given by:
\begin{equation}  
ImQ=\left\{\begin{array}{c} {\frac{4f}{\pi \sqrt{1-f^{2} +2fq} } {\bf K}\left(\frac{2\sqrt{fq} }{\sqrt{1-f^{2} +2fq} } \right),\, \, \, f<\sqrt{1+q^{2} } -q} \\ {\frac{2}{\pi } \sqrt{\frac{f}{q} } {\bf K}\left(\frac{\sqrt{1-f^{2} +2fq} }{2\sqrt{fq} } \right),\, \, \, \sqrt{1+q^{2} } -q<f<\sqrt{1+q^{2} } +q} \\ {0,\, \, \, f>\sqrt{1+q^{2} } +q} \end{array}\right.  
\label{33}
\end{equation} 
and
\begin{equation}  
ReQ=\left\{\begin{array}{c} {0,\, \, \, \, f<\sqrt{1+q^{2} } -q} \\ {\frac{2}{\pi } \sqrt{\frac{f}{q} } {\bf K}\left(\frac{\sqrt{f^{2} +2fq-1} }{2\sqrt{fq} } \right),\, \, \, \, \sqrt{1+q^{2} } -q<f<\sqrt{1+q^{2} } +q} \\ {\frac{4f}{\pi \sqrt{f^{2} +2fq-1} } {\bf K}\left(\frac{2\sqrt{fq} }{\sqrt{f^{2} +2fq-1} } \right),\, \, \, \, f>\sqrt{1+q^{2} } +q} \end{array}\right.                  
\label{34}
\end{equation} 
\noindent Here ${\bf K}$ is the complete elliptic integral of the first kind. Since this kernel, having the weak logarithmic singularity, is complex, the boson wave function $\chi (q)$ is complex also. Partition of $\chi (q)$  into the imaginary and real parts is, in a sense, arbitrary because the phase transformation of the wave function is possible,
$\chi (q)\to \chi (q)e^{i\phi}$. 
\par It is easy to see that the kernel \eqref{33}-\eqref{34} is non-Fredholm one as well. Therefore we can expect that the spectrum of the characteristic numbers $\left\{g\right\}$ of the kernel can be continuous. Accordingly, the continuous spectrum of the massless boson energy, $g$, will be corresponded to the continuous spectrum of the eigenfunctions $\chi (q,g)$.  
\subsection{Numerical procedure}
The studies of the Fredholm equation with non-Fredholm kernels are extremely rare in the literature because the search for its solutions is highly problematic. In our case, there is the normalizing condition of the kernel eigenfunctions, 
$2\pi \int_{0}^{\infty}|\chi(q,g)|^{2}qdq=1$, that alleviates the problem. Hence, $\chi(q\to \infty,g)\to 0$. 
Assume that $\chi (q,g)$ goes to zero fast enough so that Eq. \eqref{30} can be reduced to the form:
\begin{equation}  
\chi (q)=\frac{\alpha}{g} \int_{0}^{f_{0}}Q(q,f)\chi (f)df ,                                                        
\label{35}
\end{equation} 
where the value of $f_{0}(g)$ in units of ${\mathchar'26\mkern-10mu\lambda}_{e}^{-1}$ depends on the boson kinetic energy, 
and $q\in [0,f_{0}]$. 
\par The kernel $Q(q,f)$ given by \eqref{33}-\eqref{34}, is replaced by the two $N\times N$ matrices: 
\[Q_{ij}^{(r)} +iQ_{ij}^{(i)} =\frac{f_{0} }{N-1} Q(q_{i} ,f_{j}), \] 
which is denoted as ${\bf Q}^{(r)} +i{\bf Q}^{(i)}$. Here $N$ is the partition number of the interval $[0,f_{0}]$. The function 
$\chi (q)$ is replaced by the two $N$-dimensional vectors $\chi (q)\Rightarrow {\bm \chi}^{(r)} +i{\bm \chi}^{(i)}$. Then Eq. \eqref{35} is reduced to two related linear equations:
\begin{equation}  
\left(\frac{g}{\alpha} {{\bf I}}-{\bf Q}^{(r)}\right){\bm \chi}^{(r)}=-{\bf Q}^{(i)}{\bm \chi}^{(i)},
\label{36}
\end{equation} 
\begin{equation}  
\left(\frac{g}{\alpha}{{\bf I}}-{\bf Q}^{(r)}\right){\bm \chi}^{(i)}={\bf Q}^{(i)}{\bm \chi}^{(r)}. 
\label{37}
\end{equation} 
Here ${\bf I}$ is the unit $N\times N$ matrix.
\par From \eqref{36}-\eqref{37} we obtain the homogeneous system of linear \textit{N}-equations for the vector${\bm \chi}^{(r)}$:
\begin{equation}  
{{\bf A}}{\bm \chi}^{(r)}=0,                                                                
\label{38}
\end{equation} 
where the ${\bf A}$ matrix is:
\begin{equation}  
{\bf A}=\frac{g}{\alpha }{\bf I}-{\bf Q}^{(r)}+{\bf Q}^{(i)}\left(\frac{g}{\alpha}{\bf I}-{\bf Q}^{(r)}\right)^{-1}{\bf Q}^{(i)}  
\label{39}
\end{equation} 
Here $\left(\frac{g}{\alpha}{\bf I}-{\bf Q}^{(r)}\right)^{-1}$ means the inverse of 
$\left(\frac{g}{\alpha}{\bf I} -{\bf Q}^{(r)} \right)$.
\par In contrast to the Fredholm procedure, to solve the equation \eqref{38} with the definition \eqref{39} it is required to introduce a boundary condition. Taking into account the normalization of the wave functions, it is sufficient to put: 
\begin{equation}  
Re\chi (f_{0})=\delta.                                                           
\label{40}
\end{equation} 
Here $\delta$ is a small value. Typically this value is assumed to be equal to $\delta =10^{-7}$.  This value does not matter because of the subsequent normalization of the wave function.
\par Using the boundary condition \eqref{40}, from Eq. \eqref{38} we find the vector ${\bm \chi}^{(r)}$ which is the real part of the boson wave function. Then, from the equation \eqref{37} rewritten as:
\begin{equation}  
{\bm \chi}^{(i)} =\left(\frac{g}{\alpha } {\bf I} -{\bf Q}^{(r)} \right)^{-1} {\bf Q}^{(i)}{\bm \chi}^{(r)}, 
\label{41}
\end{equation} 
the vector $\bm \chi ^{(i)}$ representing the imaginary part of the massless boson wave function, is determined.  
\par After normalization of the transverse wave function ${\bm\chi}^{(r)}+i{\bm \chi}^{(i)}$ that is served as the initial approximation, the equation \eqref{36} is represented as
\begin{equation} 
{\bm \chi}^{(r)} =-\left(\frac{g}{\alpha } {\bf I} -{\bf Q}^{(r)} \right)^{-1}{\bf Q}^{(i)}{\bm \chi}^{(i)}, 
\label{42} 
\end{equation} 
and the system of two interrelated equations \eqref{41} and \eqref{42} was  solved by the  iterative method. The number of iterations was of the order of a few hundred to provide the convergent solution.
\par Then we find the average momentum of the transverse motion of the strongly coupled electron-positron pair:
\begin{equation} 
q_{av} {\mathchar'26\mkern-10mu\lambda} _{e} =2\pi \int _{0}^{f_{0}}|\chi (q,g)|^{2} q^{2} dq. 
\label{43} 
\end{equation} 
\par Finally, the transverse wave function in the coordinate representation is derived: 
($x=\rho_{34} /{\mathchar'26\mkern-10mu\lambda}_{e}$):
\begin{equation} 
\chi (x)=\int_{0}^{f_{0}}qJ_{0}(xq)\chi(q)dq  
\label{44} 
\end{equation} 
and the average transverse radius of the massless boson wave function is calculated:
\begin{equation} 
\rho_{av} /{\mathchar'26\mkern-10mu\lambda}_{e} =2\pi \int_{0}^{f_{0}}|\chi (x,g)|^{2} x^{2} dx. 
\label{45} 
\end{equation} 
Substituting the normalized function $\chi (q)$ into \eqref{44}, the wave function $\chi (x)$ obtained numerically was always normalized.
\par For all results presented below, $N=4501$ was used. To obtain reproducible results, the upper limit of integration $f_{0}$ (in units of ${\mathchar'26\mkern-10mu\lambda}_{e}^{-1}$) has a bottom restriction which depends on the boson kinetic energy. 
Therefore, $f_{0}$ was chosen separately for each energy $g$.
\subsection{Numerical results for the transverse wave functions}
Fig. 1 shows the boson energy dependencies of the integral characteristics \eqref{43} and \eqref{45} for the massless boson states. With increasing the energy the average momentum of the transverse motion of the coupled electron-positron pair increases, and the average transverse radius of the boson wave function decreases. In the low energy region $129$eV$\le g\le 511$eV, with decreasing the energy the average momentum $q_{av}$ approaches monotonically to ${\mathchar'26\mkern-10mu\lambda}_{e}^{-1}$, and the average transverse radius increases sharply, as shown in inset in Fig. 1. In a narrower region $129$eV$\le g\le 220$eV the $g$dependence of $\rho_{av}$ is close to logarithmic, $\rho_{av}\propto -\log(g)$. There is reason to hope that this logarithmic divergence persists to the limit $g\to 0$. It is consistent with the fact that the massless particles cannot be at rest. 
\begin{figure}[ht]
\centering
\includegraphics[width=8.5cm]{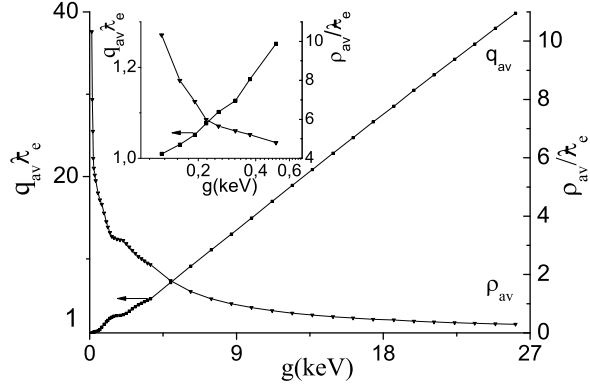}
\caption{
The average transverse momentum $q_{av}$ \eqref{43} and the average transverse radius $\rho _{av}$ \eqref{45} as  
functions of the massless boson kinetic energy. Insert:  the same in the low energy region.
\label{f1}}
\end{figure}
\par In the region with the sharp change of $\rho_{av}$, the boson wave functions in the momentum and coordinate representations are shown in Fig. 2. The boson energy is equal to $g=160$eV (the dimensionless value of $g=0.043\alpha$). Above we pointed out the arbitrariness of the choice of the imaginary and real parts of $\chi$. Therefore we do not introduce the corresponding notations for the curves shown. Obviously that one of the presented components of the wave function substantially dominates the other. However, this is one of the possible representations of $\chi(q)$, since the phase transformation of the wave function can 
change this situation.
\begin{figure}[ht]
\centering
\includegraphics[width=8.5cm]{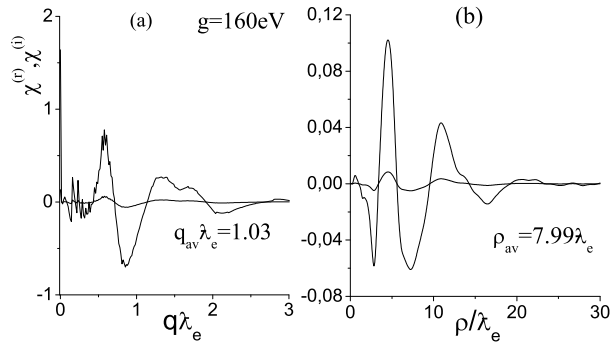}
\caption{
The momentum wave function (a) and the coordinate wave function (b) for the massless boson energy 160eV. 
The parameter $f_{0}$=65.   
\label{f2}}
\end{figure}
\par For low kinetic energies, the probability density in the momentum space has a maximum value at $q=0$ and then abruptly decreases at small $q$. With decreasing energy, the height of this maximum increases sharply. This change for small boson energies
does not allow us  to calculate accurately the coordinate wave function at $\rho=0$, since the integrand function in \eqref{44} vanishes when $q=0$. However, this does not affect the average transverse radius of the massless boson wave function. 
It was found $q_{av} =1.03{\mathchar'26\mkern-10mu\lambda}_{e}^{-1}$ and $\rho_{av} =7.99{\mathchar'26\mkern-10mu\lambda}_{e}$ for the energy $g=160$eV.
\par Computational noise on the curves presented in Fig.2a, correlates with the step of the finite difference grid 
$\Delta q=f_{0}(g)/(N-1)$. With decreasing the energy $g<160$eV the region of this noise becomes more extended. After the transformation \eqref{44} these noise features vanish, as shown in Fig. 2b. 
\par The reducing of $\rho_{av}$ with $g$ means the wave-function compression in the plane perpendicular to the boson moment $\bf g$. As a result, the electron and positron become closer to each other in the $\bm rho$-space. This wave-function compression with increasing the boson energy is especially pronounced in the region of low energies, as shown in Fig. 1. Fig. 3 shows both the momentum and coordinate wave functions for the kinetic energy $g=626$eV (the dimensionless value of $g=0.168\alpha$). The wave function in the momentum space presented in Fig. 3a, is more extended as compared with that for $g=160$keV (see Fig. 2a). Consequently, the average transverse momentum increases to the value of  $q_{av} =1.40{\mathchar'26\mkern-10mu\lambda}_{e}^{-1}$. The probability density at $q=0$ decreased significantly in comparison with that presented in Fig. 2a.
\begin{figure}[ht]
\centering
\includegraphics[width=8.5cm]{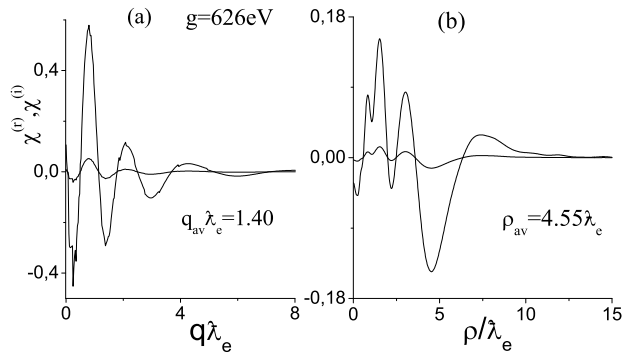}
\caption{
The same as in Fig. 2 except for the boson energy equal to 626eV. The parameter $f_{0}$=100. 
\label{f3}}
\end{figure}
\par Comparing the data in Fig. 2b and Fig. 3b, one can see that the spatial localization region of the coordinate wave function 
has significantly decreased. As a result, the average transverse radius of the boson wave function is changed from 
$\rho _{av} =7.99{\mathchar'26\mkern-10mu\lambda}_{e}$ at $g=160$eV to the value 
$\rho _{av} =4.55{\mathchar'26\mkern-10mu\lambda}_{e}$ at $g=626$eV. 
\par The transverse compression of the boson wave function with increasing the boson energy is an important feature that will be traced in subsequent results.
\par According to Fig. 1, the region 1keV$<g<$2.3keV can be considered as a transition region, in which 
$q_{av}{\mathchar'26\mkern-10mu\lambda}_{e}\simeq \rho _{av}{\mathchar'26\mkern-10mu\lambda}_{e}^{-1}$.
Beyond the region $\rho_{av}$ decreases monotonically with increasing the boson energy, and the dependence of $q_{av}(g)$ becomes close to linear. 
\begin{figure}[ht]
\centering
\includegraphics[width=8.5cm]{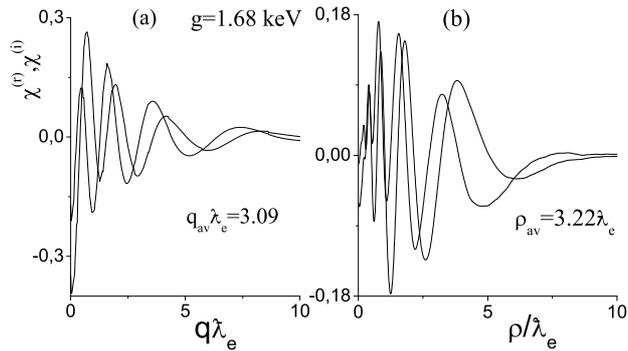}
\caption{
The same as in Fig. 2 except for the massless boson energy equal to 1.68 keV. The parameter $f_{0}$=130. 
\label{f4}}
\end{figure}
\par In this transition region, the massless boson wave functions in the momentum and coordinate spaces are presented in Fig. 4 for the energy $g=1.68$keV. Comparing the data in Fig. 3 and Fig. 4, one can conclude that with increasing the boson energy, oscillations of the wave functions in both the momentum and coordinate spaces are enhanced. The extension of the momentum wave function increases (Fig. 4a) and, accordingly, the  localization region of the coordinate wave function decreases (Fig. 4b). 
\par Note that in the coordinate representation, with the greatest probability density the electron and positron are still at a finite distance from each other. This distance is approximately equal to $1.25{\mathchar'26\mkern-10mu\lambda}_{e}$, as shown 
in Fig. 4b. With increasing the energy of the composite boson, the mean distance between the electron and positron become shorter, but the transverse compression of the boson wave function $\chi(\rho ,g)$ tends to slow down. It is demonstrated in 
Fig. 5, where the momentum and coordinate wave functions of the composite boson are presented for the kinetic energy $g=13.65$keV (the dimensionless value of $g=3.66\alpha$).  
\begin{figure}[ht]
\centering
\includegraphics[width=8.5cm]{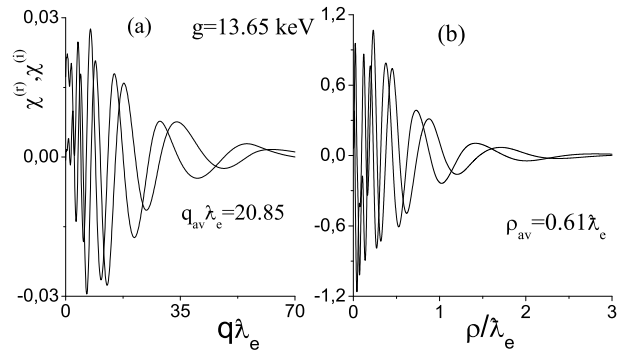}
\caption{
The same as in Fig. 2 except for the boson energy equal to 13.65 keV. The parameter $f_{0}$=467. 
\label{f5}}
\end{figure}
\par The wave function in the momentum space presented in Fig. 5a, is more extended as compared with that shown in Fig. 4a. Consequently, the transverse momentum is increased from $q_{av} =3.09{\mathchar'26\mkern-10mu\lambda}_{e}^{-1}$ 
at $g=1.68$keV to $q_{av} =20.85{\mathchar'26\mkern-10mu\lambda}_{e}^{-1}$ at $g=13.65$keV. The spatial compression of the 
coordinate wave function is very evident when comparing the data in Fig. 4b and Fig. 5b. 
The transverse size of the wave function is decreased from $\rho_{av} =3.22{\mathchar'26\mkern-10mu\lambda}_{e}$ to 
$\rho_{av} =0.61{\mathchar'26\mkern-10mu\lambda}_{e}$. 
\par Thus, the found changes in the wave function with increasing the boson energy can be summarized as follows: 
the extension of the momentum wave function leading to the average transverse momentum increase, and the compression of the
coordinate wave function accompanying with the decrease of the average transverse distance between the electron and positron.  The wave function compression has a feature that can be seen in Fig. 6 where the boson state corresponding to the kinetic energy $g=24.87$keV (the dimensionless value of$g=6.67\alpha $), is presented. 
\begin{figure}[ht]
\centering
\includegraphics[width=8.5cm]{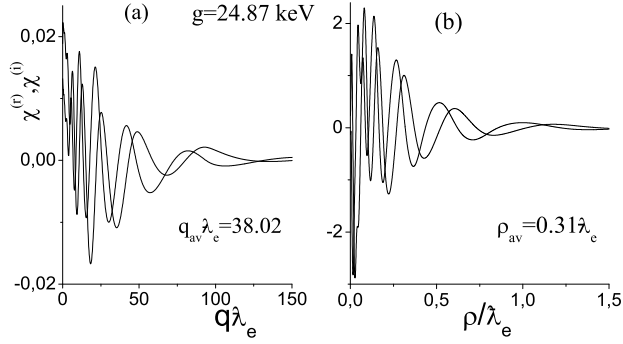}
\caption{
The massless boson wave function in the momentum space (a) and the coordinate space (b)  
for the kinetic energy 24.87 keV. The parameter $f_{0}$=767. 
\label{f6}}
\end{figure}
\par The transverse momentum wave function (Fig. 6a) is very extended, and the average transverse momentum  
$q_{av} =38.02{\mathchar'26\mkern-10mu\lambda}_{e}^{-1}$. The wave function in coordinate space is presented in Fig. 6b.
The average transverse radius between the electron and positron $\rho_{av} =0.31{\mathchar'26\mkern-10mu\lambda}_{e}$. 
It can be seen  that with the greatest probability density the electron and positron are at low distances from each other. The probability density is maximal in the region $\rho<0.1{\mathchar'26\mkern-10mu\lambda}_{e}$.   
\par So, with increasing the composite boson energy the electron and positron become more and more closer to each other. 
In this situation, one can expect a small cross-section for the interaction of the massless composite bosons,
formed by the coupled electron-positron pairs, with, for example, photons and fermions. 
\section{The $2\gamma$ angular correlation spectrum}
It was shown above that the consideration of the electron and positron as independent particles leads to the appearance of the
branch of the massless composite bosons formed by the coupled electron-positron pairs with the coupling equal to the fine structure constant. In this Section, the reaction \eqref{1} is theoretically investigated. Then, an experiment that will establish 
whether there is the conventional annihilation of singlet electron-positron pair with the two-photon emission or the investigated reaction, products of which are the three particles, is proposed.
\subsection{The initial and final states of the reaction \eqref{1}}
First, we need the massless boson state for an arbitrary direction of the boson momentum. When the momentum, $\bf g$, is directed along the \textit{z}-axis, the boson wave function is given by \eqref{10} with 
$\varphi_{1,2}({\bm \rho},{\bf g})=\chi_{1,2}({\bm \rho},{\bf g})v_{1,2}$, where, according to \eqref{18}, the bispinor parts have the form: 
\begin{displaymath} 
v_{1} =\frac{1}{\sqrt{2}} \left[\left(\begin{array}{c}{1} \\ {0} \\ {0} \\ {0} \end{array}\right)_{-} \left(\begin{array}{c} {0} \\ {0} \\ {0} \\ {1} \end{array}\right)_{+}+\left(\begin{array}{c} {0} \\ {0} \\ {0} \\ {1} \end{array}\right)_{-}\left(
\begin{array}{c} {1} \\ {0} \\ {0} \\ {0} \end{array}\right)_{+} \right],\, \, \, \, 
\end{displaymath} 
\begin{equation}\label{46} 
v_{2} =\frac{1}{\sqrt{2}} \left[\left(\begin{array}{c} {0} \\ {1} \\ {0} \\ {0} \end{array}\right)_{-} \left(\begin{array}{c} {0} \\ {0} \\ {1} \\ {0} \end{array}\right)_{+} +\left(\begin{array}{c} {0} \\ {0} \\ {1} \\ {0} \end{array}\right)_{-} \left(
\begin{array}{c} {0} \\ {1} \\ {0} \\ {0} \end{array}\right)_{+} \right].              
\end{equation} 
Here the bispinor marked with the minus sign refers to the electron, and the bispinor labeled with the plus sign refers to the positron. The bispinors $v_{1,2}$ for which the wave functions of the transverse motion $\chi_{1,2}({\bm \rho},g)$ are the same
($\chi_{1}=\chi_{2}=\chi$), are defined only the projection of the total spin on the ${\bf g}$ direction, which is equal to zero, 
$(\Sigma_{z-}+\Sigma_{z+} )v_{1,2}=0$. Therefore, we can say that in the states of massless composite boson the helicities of the electron and positron are opposite.
\par In the case of boson motion along $x$-axis boson (${\bf g}//x$-axis), it is easy to obtain that the bispinor parts in the state \eqref{10} are of the form:
\begin{displaymath}
v_{1}=\frac{1}{\sqrt{8}} \left[\left(\begin{array}{c} {1} \\ {1} \\ {0} \\ {0} \end{array}\right)_{-} \left(\begin{array}{c} {0} \\ {0} \\ {1} \\ {-1} \end{array}\right)_{+} +\left(\begin{array}{c} {0} \\ {0} \\ {1} \\ {-1} \end{array}\right)_{-} \left(
\begin{array}{c} {1} \\ {1} \\ {0} \\ {0} \end{array}\right)_{+} \right],\, \, \, \, 
\end{displaymath}
\begin{equation}\label{47}  
v_{2} =\frac{1}{\sqrt{8}} \left[\left(\begin{array}{c} {1} \\ {-1} \\ {0} \\ {0} \end{array}\right)_{-} \left(\begin{array}{c} {0} \\ {0} \\ {1} \\ {1} \end{array}\right)_{+} +\left(\begin{array}{c} {0} \\ {0} \\ {1} \\ {1} \end{array}\right)_{-} \left(
\begin{array}{c} {1} \\ {-1} \\ {0} \\ {0} \end{array}\right)_{+} \right].             
\end{equation} 
For the function \eqref{47} we have $(\Sigma_{x-}+\Sigma_{x+})v_{1,2} =0$. It seems obvious the form of these functions when 
${\bf g}//y$-axis.  
\par Now consider an arbitrary direction of the momentum boson. Let ${\bf r}_{\bot}$ is an arbitrary radius vector lying in the plane perpendicular to the vector ${\bf g}$ (${\bf gr}_{\bot} =0$). Then the wave function \eqref{10} takes the form:
\begin{equation} \label{48} 
\psi_{1,2} =\chi({\bf r}_{\bot},g)v_{1,2}({\bf i}_{\bf g})\exp(i{\bf gr}_{//}-igt), 
\end{equation} 
\noindent where ${\bf r}_{//} ={\bf r}-{\bf r}_{\bot}$ is the component of the radius vector ${\bf r}$ which is collinear to 
${\bf g}$, $v_{1,2}$ are the bispinor parts which can depend only on the angles of the unit vector ${\bf i}_{\bf g} ={\bf g}/g$. For the wave function \eqref{48} one should understand that ${\bf r}_{\bot} ={\bf r}_{1\bot} -{\bf r}_{2\bot}$, where 
${\bf gr}_{1\bot}=0$ and ${\bf gr}_{2\bot} =0$, as well as that ${\bf r}_{1//} ={\bf r}_{1} -{\bf r}_{1\bot}$ and  
${\bf r}_{2//} ={\bf r}_{2}-{\bf r}_{2\bot}$ with the condition ${\bf r}_{1//} ={\bf r}_{2//} ={\bf r}_{//} $.
\par As is well known, the three-dimensional spinors corresponding the definite helicities, have the form: 
\begin{equation} \label{49}
w_{\frac{1}{2}}=\left(\begin{array}{c} {e^{-i\varphi_{g}/2} \cos \frac{\theta_{g}}{2}} \\ {e^{i\varphi_{g}/2} \sin 
\frac{\theta_{g}}{2}}\end{array}\right), w_{-\frac{1}{2}}=\left(\begin{array}{c}{-e^{-i\varphi_{g}/2} \sin 
\frac{\theta_{g}}{2}} \\ {e^{i\varphi_{g} /2}\cos \frac{\theta_{g}}{2}} \end{array}\right),                                  
\end{equation}
\noindent where $\theta_{g}$ and $\varphi_{g}$ are the polar and azimuthal angles of the vector ${\bf i}_{\bf g}$.  
Taking into account \eqref{49}, one can see that \eqref{46} and \eqref{47} are the special cases of the following bispinor functions:
\begin{displaymath}
v_{1}=\frac{1}{\sqrt{2}}\left[\left(\begin{array}{c}{w_{\frac{1}{2}}} \\ {0} \end{array}\right)_{-} \left(\begin{array}{c} {0} \\ 
{w_{-\frac{1}{2}}} \end{array}\right)_{+}+\left(\begin{array}{c} {0} \\ {w_{-\frac{1}{2}}} \end{array}\right)_{-} \left(
\begin{array}{c}{w_{\frac{1}{2}}} \\ {0} \end{array}\right)_{+}\right],
\end{displaymath}
\begin{equation} \label{50} 
v_{2}=\frac{1}{\sqrt{2}}\left[\left(\begin{array}{c}{w_{-\frac{1}{2}}} \\ {0} \end{array}\right)_{-} \left(\begin{array}{c} {0} \\ 
{w_{\frac{1}{2}}} \end{array}\right)_{+}+\left(\begin{array}{c} {0} \\ {w_{\frac{1}{2}}} \end{array}\right)_{-} \left(
\begin{array}{c}{w_{-\frac{1}{2}}} \\ {0} \end{array}\right)_{+}\right].     
\end{equation} 
\par Now we turn to the initial state of the free electron and positron. In principle one can consider colliding non-spin-polarized particles with the equal  energy. Currently, however, the production of spin-polarized low-energy positron beams with the kinetic energy of a few electron volts became possible \cite{Pus,Lyn,Col,Zit,Gri}. It is therefore of particular interest to study the annihilation-like process \eqref{1} for the polarized beams.
\par Suppose that the electron and positron collide with the center of mass at rest (${\bf p}_{-}+{\bf p}_{+}=0$). For definiteness let the the electron momentum, ${\bf p}_{-}$, be directed along the z-axis; then the positron momentum, ${\bf p}_{+}$, - against the z-axis. The wave function of the free pair can be represented as:
\begin{equation} \label{51} 
\varphi(1,2)=u\exp(-ip_{-}x_{-}-ip_{+}x_{+}),                                               
\end{equation} 
where $p_{-} =(\varepsilon _{-},{\bf p}_{-})$ and $x_{-} =(t,{\bf r})$ are the four-vectors for the electron, 
$p_{+} =(\varepsilon _{+},{\bf p}_{+})$ and $x_{+} =(t,{\bf r}_{1} )$  are the four-vectors of the positron, and 
${\bf p}_{-} =-{\bf p}_{+} ={\bf p}$. The spin function in \eqref{51} can be defined by the $\it z$-projections of the spin for the particles in the rest system. Then, for the total spin projection $S_{z}=0$  the function $u$ can be represented as:
\begin{equation} \label{52} 
u_{S_{z} =0} =\left(\begin{array}{c} {\sqrt{\varepsilon +m}} \\ {0} \\ {\sqrt{\varepsilon -m} } \\ {0} \end{array}\right)_{-} 
\left(\begin{array}{c} {0} \\ {\sqrt{\varepsilon +m}} \\ {0} \\ {\sqrt{\varepsilon -m}} \end{array}\right)_{+}. 
\end{equation} 
\par The function \eqref{52} can be considered as the two-particle spin state of with opposite helicities for these particles. One can also say that \eqref{52} corresponds to the longitudinal polarization of the particles.
\par Free pairs of the transversely polarized particles relative to the vector $\bf p$ can be prepared as well, and it is easy to write, for example, $u_{S_{x} =0}$ for the polarization along the \textit{x}-axis. Besides, the colliding electron-positron pair with $S_{z} =1$ or $S_{x} =1$  can be experimentally obtained, and it is not difficult to write the corresponding bispinor parts for these cases.
\par Interaction for the reaction \eqref{1}, which will be discussed in Section 3.2, predetermines the choice of the photon wave function. The photon plane wave can be represented as:
\begin{equation} \label{53} 
A_{\bf k}^{(\alpha)}=\sqrt{4\pi}{\bf l}_{\bf k}^{(\alpha)}\exp(-ikx),                                                
\end{equation} 
where ${\bf l}_{\bf k}^{(\alpha )}$ is the photon polarization that can be chosen real, $kx=\omega _{k} t-{\bf kr}$, 
$\omega_{k},{\bf k}$ are the energy and wave vector of the photon, respectively. For the transverse photons  
${\bf kl}_{k}^{(\alpha)}=0$ with $\alpha =1,2$.
\subsection{Interaction for the reaction \eqref{1}}
The products of reaction \eqref{1} include two photons and the massless boson. This composite boson is formed by the strongly coupled electron-positron pair. Therefore, the radiative transition of any particle from the initial free pair (the left-hand side of \eqref{1}) to any intermediate state does not lead to the formation of the massless boson. Only the simultaneous emission of one photon by the electron and other photon by the positron and, accordingly, the simultaneous transition of both the electron and positron into the massless boson state  is the only process to occur the reaction \eqref{1}. For this reason it is necessary to determine the interaction operator for the simultaneous emission of photons by each particle of the pair.
\par The additive energy of any free pair from the beams is defined as:
\begin{equation} \label{54} 
\varepsilon_{-} +\varepsilon_{+}=\sqrt{m_{s}^{2}+({\bf p}_{-}+{\bf p}_{+})^{2}}.                                            
\end{equation} 
\noindent Here $m_{s}$ is the pair mass. Note that in \eqref{54} the sum of the particle momenta is presented. 
\par The operator of single-photon emission by the electron and positron at the same time can be obtained from \eqref{54} by introducing the canonical momenta in the presence of electromagnetic fields, ${\bf p}_{-}\to{\bf p}_{-}+e{\bf A}_{-}$ and  
${\bf p}_{+}\to{\bf p}_{+}-e{\bf A}_{+}$ (${\bf A}_{\pm }$ are the operators of the vector potentials generated by the electron and positron). Assuming $m_{s} =2m$ for the non-relativistic particles, from \eqref{54} we obtain the expression for this operator:
\begin{equation} \label{55} 
W=-\frac{e^{2}}{2m}{\bf A}_{-}{\bf A}_{+}.                                                        
\end{equation} 
\par A similar approach can be applied to positronium. The kinetic energy of the unperturbed Hamiltonian is:
\begin{displaymath}
T=\frac{{\bf p}^{2} }{m}, 
\end{displaymath}
where ${\bf p}$ is the momentum operator of the relative motion, corresponding the relative radius vector between the particles. Substituting ${\bf p}=({\bf p}_{-}-{\bf p}_{+})/2$ into this energy and making the transition to the canonical momenta of the particles, we get, up to the sign, the operator \eqref{55} for simultaneous emission of two photons by the two particles.
\par The electromagnetic interaction operator \eqref{55} is used below for the study of reaction \eqref{1}. 
\subsection{Possible reaction channels}
Now, taking into account the spin conservation, we discuss the possible reaction channels with initially free electron-positron pairs. Let the colliding pairs are in the triplet state, $S_{z} =1$. First of all, the reaction with single-photon emission
\begin{displaymath}
(e^{-}e^{+})_{S_{z} =1}\to B\gamma.                                                        
\end{displaymath}
is {\it impossible in principle}. Here $(e^{-}+e^{+} )_{S_{z} =1}$ denotes the initial free pair, $B$ is the massless boson,  and  $\gamma $ is the photon emitted.  For this reaction, the single photon is only emitted either by the electron or by the positron. Therefore, the simultaneous transition of these particles into the composite massless boson state cannot occur.
\par The reaction with emission of three photons, 
\begin{equation} \label{56} 
(e^{-}e^{+})_{S_{z} =1}\to B\gamma_{1}\gamma_{2}\gamma_{3}  
\end{equation} 
can take place. For \eqref{56}, at first one of the particles (electron or positron) emits one photon (say $\gamma_{1}$) and passes in an intermediate state, and then in a subsequent point in time, simultaneous emission of two other photons 
($\gamma_{2},\gamma_{3}$) is accompanied the simultaneous transition of these particles into the boson state $B$. 
\par In the case of prepared pairs with $S_{z} =0$ the reaction with single-photon emission is also forbidden in principle as was discussed above. 
\par The reaction with the emission of two photons:
\begin{equation} \label{57} 
(e^{-}e^{+})_{S_{z}=0}\to B\gamma_{1}\gamma_{2}.                                            
\end{equation} 
can occur due to the interaction \eqref{55}. As will be shown below, for \eqref{57} it is possible and preferably the emission of composite bosons with relatively low energy that is several orders of magnitude smaller than the electron mass. With increasing the energy of boson emitted the cross section of this reaction decreases sharply.
\par Below, we derive the cross section of the reaction \eqref{57} and study numerically the $2\gamma$ angular correlation spectra.
\subsection{Cross section of the reaction \eqref{57}}
In the center-of-mass system ${\bf p}_{-}=-{\bf p}_{+}={\bf p}$ and $\varepsilon_{-} +\varepsilon_{+} =2\varepsilon$. 
The cross-section of the reaction \eqref{57} has the form:
\begin{equation}\label{58} 
d\sigma =(2\pi )^{4} \delta^{(4)}(P_{i}-P_{f})\left|W_{if} \right|^{2} \frac{1}{4\varepsilon^{2} v_{rel}} 
\frac{d{\bf k}}{(2\pi )^{3}2\omega} \frac{d{\bf k}_{1}}{(2\pi )^{3} 2\omega_{1}} \frac{d{\bf g}}{(2\pi )^{3}2g}.
\end{equation} 
\noindent Here $\delta^{(4)}(P_{i}-P_{f})=\delta (\omega +\omega _{1}+g-2\varepsilon)\delta({\bf k}+{\bf k}_{1}+{\bf g})$, 
$v_{rel} =2v_{e}$ is the relative velocity of the particles, $v_{e} =\sqrt{2(\varepsilon-m)/m}$, $W_{if}$ is the matrix element of the operator \eqref{55} for the transition from the initial state (\textit{i}) of the free pair \eqref{51}-\eqref{52} into the final state (\textit{f}). The latter include the composite massless boson \eqref{48}-\eqref{50} and the two photons \eqref{53}, 
one of which has the energy $\omega$, the momentum ${\bf k}$ and the polarization ${\bf l}_{{\bf k}}^{(\alpha)}$ ($\alpha =1,2$) and the second photon -  $\omega_{1}$, ${\bf k}_{1}$ and ${\bf l}_{{\bf k}_{1}}^{(\beta)}$ ($\beta =1,2$), respectively. 
This matrix element is written as:
\begin{displaymath}
W_{if} =-4\pi \frac{e^{2}}{2m} \left({\bf l}_{{\bf k}}^{(\alpha)}{\bf l}_{{\bf k}_{1}}^{(\beta)} \right)
\left(v_{1,2}^{+} u_{S_{z} =0} \right)
\end{displaymath}
\begin{equation} \label{59}
\int d{\bf r} \int d{\bf r}_{1} \chi ^{*}(|{\bf r}_{\bot}-{\bf r}_{1\bot} |;g)e^{-i{\bf kr}-i{\bf k}_{1}{\bf r}_{1}-
i{\bf gr}_{//}} \left(e^{i{\bf pr}-i{\bf pr}_{1}}+e^{i{\bf pr}_{1}-i{\bf pr}} \right). 
\end{equation} 
\par On the right side \eqref{59}, the last factor in round brackets is the sum of two terms. The first contribution corresponds to the emission of photon with the wave vector ${\bf k}$ by the electron, and in the second term the electron emits photon with the wave vector ${\bf k}_{1}$. The second of two photons is emitted simultaneously by the positron.
\par Now we calculate the spatial integrals in \eqref{59}. For this purpose we introduce the notations:  
${\bf r}_{//} =({\bf ri}_{\bf g}){\bf i}_{\bf g}$ is the component of the vector ${\bf r}$ directed along the vector
${\bf g}$, ${\bf r}_{1//} =({\bf r}_{1} {\bf i}_{\bf g}){\bf i}_{\bf g}$ is the same for ${\bf r}_{1}$. 
Then ${\bf r}_{\bot} ={\bf r}-({\bf ri}_{{\bf g}} ){\bf i}_{{\bf g}}$ is the component of the vector ${\bf r}$ that is perpendicular to the vector ${\bf g}$, and ${\bf r}_{1\bot} ={\bf r}_{1}-({\bf r}_{1}{\bf i}_{\bf g}){\bf i}_{\bf g}$ is the 
same for the vector ${\bf r}_{1}$. We have $d{\bf r}=d{\bf r}_{\bot} dr_{//}$ and $d{\bf r}_{1} =d{\bf r}_{1\bot}dr_{1//}$, where $r_{//} ={\bf ri}_{\bf g}$ and $r_{1//} ={\bf r}_{1}{\bf  i}_{\bf g}$. 
\par As noted above, when the boson momentum is directed along \textit{z}-axis, the \textit{z}-components of the radius-vectors of particles coincide, $z_{1} =z_{2} =z$. Now we need to take into account that $r_{//} =r_{1//}$ for the composite boson and pass to the new variables of integration. Introducing ${\bf r}_{\bot} ={\bf R}+{\bm \rho} /2$ and ${\bf r}_{1\bot} ={\bf R}-{\bm \rho} /2$, where ${\bm \rho} ={\bf r}_{\bot} -{\bf r}_{1\bot}$ is the two-dimensional relative vector between the electron and positron, as a result we obtain:
\begin{displaymath}
\int d{\bf r}\int d{\bf r}_{1}\chi^{*}(|{\bf r}_{\bot}-{\bf r}_{1\bot}|;g) e^{-i{\bf kr}-i{\bf k}_{1}{\bf r}_{1}-i{\bf gr}_{//}}  
\left(e^{i{\bf pr}-i{\bf pr}_{1}}+e^{i{\bf pr}_{1} -i{\bf pr}}\right)
\end{displaymath}
\begin{equation} \label{60} 
=\chi ^{*}(|{\bf k}_{\bot }-{\bf p}_{\bot} |;g)+\chi ^{*}(|{\bf k}_{\bot}+{\bf p}_{\bot}|;g)
\end{equation} 
\noindent multiplied by $(2\pi )^{3}\delta({\bf k}+{\bf k}_{1}+{\bf g})$. We do not write the latter, because it is already included in \eqref{58}. In \eqref{60} the Fourier transform of the wave function of the transverse motion of the coupled pair in the massless composite boson state is used:
\begin{displaymath}
\chi^{*}(|{\bf k}_{\bot}+{\bf p}_{\bot}|;g)=\int \chi ^{*}(\rho ;g)e^{-i({\bf k}+{\bf p}){\bm \rho}} d{\bm \rho}.  
\end{displaymath}
\noindent Here ${\bf k}_{\bot}$ is the component of the photon wave vector ${\bf k}$ which is perpendicular to the boson momentum ${\bf g}$, 
\begin{equation} \label{61} 
{\bf k}_{\bot} ={\bf k}-\frac{({\bf kg}){\bf g}}{g^{2}},                                                   
\end{equation} 
\noindent ${\bf p}_{\bot}$ is the component of the free electron momentum perpendicular to ${\bf g}$,
\begin{equation} \label{62} 
{\bf p}_{\bot } ={\bf p}-\frac{({\bf pg}){\bf g}}{g^{2}}.                                                   
\end{equation} 
\noindent According to the momentum conservation, ${\bf k}_{\bot} +{\bf k}_{1\bot} =0$, where ${\bf k}_{1\bot}$ is the corresponding component of the wave vector of the second emitted photon.
\par Note that $\chi ^{*}(|{\bf k}_{\bot}+{\bf p}_{\bot}|;g)$ has the dimension of length.
\par Substituting \eqref{59}-\eqref{60} in \eqref{58}, the cross-section is given by:
\begin{displaymath}
d\sigma=2^{-8}\pi^{-3}r_{e}^{2}\frac{c}{v_{rel}}\sum_{\alpha,\beta}\left|{\bf l}_{{\bf k}}^{(\alpha)}
{\bf l}_{{\bf k}_{1}}^{(\beta)} \right|^{2} \frac{1}{\varepsilon ^{2}}\left|v_{1,2}^{+} u_{S_{z} =0}\right|^{2} 
\left|\chi (|{\bf k}_{\bot}-{\bf p}_{\bot }|;g)+\chi (|{\bf k}_{\bot }+{\bf p}_{\bot} |;g)\right|^{2} * 
\end{displaymath}
\begin{equation}\label{63} 
\delta(k+k_{1} +g-2k_{e})\delta ({\bf k}_{1} +{\bf k}+{\bf g})\frac{d{\bf k}}{k} \frac{d{\bf k}_{1}}{k_{1}} 
\frac{d{\bf g}}{g},
\end{equation} 
\noindent where $r_{e} =e^{2}/m$ is the electromagnetic radius of the electron and 
$k_{e} ={\mathchar'26\mkern-10mu\lambda} _{e}^{-1} *\varepsilon /m$,  ${\mathchar'26\mkern-10mu\lambda}_{e}$ is the Compton wavelength of the electron.
\subsection{Transformation of Eq. \eqref{63}}
In \eqref{63} we have the summation over photon polarizations:
\begin{displaymath}
S=\sum_{\alpha,\beta}\left|{\bf l}_{{\bf k}}^{(\alpha )}{\bf l}_{{\bf k}_{1} }^{(\beta)}\right|^{2}  
\end{displaymath}
\par Because $l_{{\bf k}i}^{(\alpha)}l_{{\bf k}j}^{(\alpha)}=0$ for $i,j=x,y,z$ and $i\ne j$, the sum is reduced to:
\begin{displaymath}
S=\sum_{\alpha,\beta}\sum_{i=x,y,z} l_{{\bf k}i}^{(\alpha)^{2}} l_{{\bf k}_{1}i}^{(\beta )^{2}}
\end{displaymath}
\par Since the polarizations are normalized to unity, we can use 
$l_{{\bf k}z}^{(\alpha )^{2}} =1-l_{{\bf k}x}^{(\alpha )^{2}}-l_{{\bf k}y}^{(\alpha )^{2}}$. 
Then, taking into account $\sum _{\alpha =1,2}l_{kx}^{(\alpha )^{2} } =1 $, finally we obtain:
\begin{equation}\label{64} 
\sum_{\alpha,\beta}\left|{\bf l}_{{\bf k}}^{(\alpha)}{\bf l}_{{\bf k}_{1}}^{(\beta )} 
\right|^{2}=2.                                                       
\end{equation} 
\par In \eqref{63} the factor
\begin{displaymath}
\frac{1}{\varepsilon ^{2}}\left|v_{1,2}^{+}u_{S_{z} =0} \right|^{2}  
\end{displaymath}
can be considered as the overlap of bispinor functions.  Using \eqref{50} and \eqref{52}, we find:
\begin{equation} \label{65} 
\frac{1}{\varepsilon ^{2} } \left|v_{1,2}^{+} u_{S_{z} =0} \right|^{2} =
\frac{p^{2} }{2\varepsilon ^{2} } \cos ^{2} \theta _{g} .                                               
\end{equation} 
\par Substituting \eqref{64} and \eqref{65} into \eqref{63}, in the non-relativistic limit ($\varepsilon \cong m$, $p=mv_{e}$) 
we obtain:
\begin{displaymath}
d\sigma =2^{-9}\pi^{-3} r_{e}^{2} \frac{v_{e} }{c} \cos^{2} \theta_{g} \left|\chi(|{\bf k}_{\bot}-{\bf p}_{\bot}|;g)+
\chi(|{\bf k}_{\bot}+{\bf p}_{\bot}|;g)\right|^{2}  
\end{displaymath}
\begin{equation} \label{66} 
 \delta(k+k_{1}+g-2k_{e})\delta({\bf k}_{1}+{\bf k}+{\bf g})\frac{d{\bf k}}{k} \frac{d{\bf k}_{1}}{ k_{1}} 
\frac{d{\bf g}}{g}.                          
\end{equation} 
\par According to the momentum conservation of the reaction products, ${\bf g}=-({\bf k}+{\bf k}_{1})$ and, hence, 
\begin{displaymath}
\cos \theta _{g} =-\frac{k\cos \theta_{k}+k_{1} \cos \theta_{k_{1}}}{|{\bf k}+{\bf k}_{1}|}, 
\end{displaymath}
where $\theta_{k}$ and $\theta_{k_{1}}$ are polar angles of the photon wave vectors ${\bf k}$  and ${\bf k}_{1}$, respectively. 
\par Integrating over the massless boson momentum ${\bf g}$, the cross section \eqref{66} takes the form:
\begin{displaymath}
d\sigma =2^{-9} \pi ^{-3} r_{e}^{2} \frac{v_{e}}{c} \frac{(k\cos \theta_{k} +k_{1} \cos \theta_{k_{1}})^{2}}
{(2k_{e}-k-k_{1})^{3}} 
\end{displaymath}
\begin{displaymath}
\left|\chi (|{\bf k}_{\bot }-{\bf p}_{\bot } |;2k_{e}-k-k_{1} )+
\chi (|{\bf k}_{\bot } +{\bf p}_{\bot } |;2k_{e}-k-k_{1} )\right|^{2}
\end{displaymath}
\begin{equation} \label{67} 
\delta (k+k_{1} +|{\bf k}_{1} +{\bf k}|-2k_{e} )kk_{1} dkdk_{1} d\Omega_{{\bf k}} d\Omega_{{\bf k}_{1}}.
\end{equation} 
\noindent Here $d\Omega_{{\bf k}} $ and $d\Omega_{{\bf k}_{1}}$ are the solid angle elements of the photon wave vectors. 
\par Now it is easy to carry out the integration over the energy of one of the two photons, for example, over $k_{1}$. As a result, from \eqref{67} we obtain:
\begin{displaymath}
\frac{\partial ^{2} \sigma}{\partial \Omega_{{\bf k}} \partial \Omega_{{\bf k}_{1}}} =
2^{-8}\pi ^{-3} r_{e}^{2} \frac{v_{e}}{c} \int_{0}^{k_{e}}\frac{k_{e} k(k_{e}-k)dk}{(2k_{e}^{2}-k(2k_{e}-k)
(1-\cos \vartheta ))^{2}} 
\end{displaymath}
\begin{equation} \label{68} 
\left(k\cos \theta_{k}+\frac{2k_{e}(k_{e}-k)}{2k_{e}-k(1-\cos \vartheta)} \cos \theta_{k_{1}} \right)^{2}    
\left|\chi (|{\bf k}_{\bot}-{\bf p}_{\bot} |;g)+\chi (|{\bf k}_{\bot }+{\bf p}_{\bot} |;g)\right|^{2}.  
\end{equation} 
Here we use the following notations: $g$ is the kinetic energy of the massless composite boson emitted,
\begin{equation} \label{69} 
g=\frac{2k_{e}^{2}-k(2k_{e}-k)(1-\cos \vartheta)}{2k_{e}-k(1-\cos \vartheta )},
\end{equation} 
\noindent $d\Omega_{{\bf k}} =\sin \theta_{k} d\theta_{k} d\varphi_{k}$, $d\Omega_{{\bf k}_{1}}=\sin \theta_{k_{1}} 
d\theta_{k_{1}} d\varphi_{k_{1}}$,  $\vartheta$ is the angle between the wave vectors ${\bf k}$ and ${\bf k}_{1}$ 
(${\bf kk}_{1}=kk_{1} \cos \vartheta$): 
\begin{displaymath}
\cos \vartheta =\cos \theta_{k}\cos \theta_{k_{1}}+\sin \theta_{k} \sin \theta_{k_{1}}\cos (\varphi_{k}-\varphi_{k_{1}}), 
\end{displaymath}
\noindent $\varphi_{k}$ and $\varphi_{k_{1}}$ are the azimuthal angles of the  wave vectors $\bf k$ and ${\bf k}_{1}$. 
\par Then, according to (61),     
\begin{equation} \label{70} 
{\bf k}_{\bot}={\bf k}-\frac{({\bf k}({\bf k}+{\bf k}_{1}))}{|{\bf k}+{\bf k}_{1} |^{2} }*({\bf k}+{\bf k}_{1}) 
\end{equation} 
\noindent and its modulus depends only on $k$ and the angle $\vartheta$,
\begin{equation} \label{71} 
k_{\bot} =\frac{2kk_{e} (k_{e}-k)\sqrt{1-\cos ^{2} \vartheta}}{2k_{e}^{2}-k(2k_{e}-k)(1-\cos \vartheta)}.
\end{equation} 
\par In accordance with (62),
\begin{equation} \label{72} 
{\bf p}_{\bot}={\bf p}-\frac{({\bf p}({\bf k}+{\bf k}_{1}))}{|{\bf k}+{\bf k}_{1} |^{2}}*({\bf k}+{\bf k}_{1} ) 
\end{equation} 
and, because ${\bf p}$ is parallel the \textit{z}-axis, the modulus of this value is 
\begin{equation} \label{73} 
p_{\bot} =p\left|\frac{{\bf k}_{\rho}+{\bf k}_{\rho 1}}{{\bf k}+{\bf k}_{1}} \right|.
\end{equation}  
\subsection {The observables}
Consider colliding two bunches of electrons and positrons with flux densities ${\rm I}_{-}$ and ${\rm I}_{+} $ that are moving towards each other.  Let ${\rm I}_{-}={\rm I}_{+}={\rm I}$, the spatial lengths of bunches $L$ and their cross-sectional areas 
$S$ are  identical. Then the number of coincidence events per time unit that due to the reaction \eqref{57} one photon is detected in the small element of solid angle $\Delta\Omega _{\bf k}$, and  the second - in $\Delta\Omega _{{\bf k}_{1} }$ has the form:
\begin{equation} \label{74} 
\frac{\Delta N}{\Delta t}=\zeta(\cos \vartheta) LS{\rm I}^{2} \Delta\Omega_{\bf k} \Delta\Omega_{{\bf k}_{1}},
\end{equation} 
where the small solid angles are determined by the angular resolution of an experimental setup, and, according to \eqref{68},  
\begin{displaymath}
\zeta(\cos \vartheta)=2^{-8}\pi ^{-3}\frac{r_{e}^{2}}{c}\int_{0}^{k_{e}}\frac{k_{e} k(k_{e}-k)dk}{(2k_{e}^{2}-k(2k_{e}-k) 
(1-\cos \vartheta))^{2}} 
\end{displaymath}
\begin{equation} \label{75} 
\left(k\cos \theta_{k}+\frac{2k_{e} (k_{e} -k)}{2k_{e}-k(1-\cos \vartheta )}\cos \theta_{k_{1}} \right)^{2}    
\left|\chi (|{\bf k}_{\bot}-{\bf p}_{\bot} |;g)+
\chi (|{\bf k}_{\bot }+{\bf p}_{\bot} |;g)\right|^{2}.  
\end{equation} 
That is, precisely the value \eqref{75} that has dimension $M\times S$, can be measured in experiments.  
\par Now we carry out the analysis of the observables \eqref{68} and \eqref{75}. If both of these values would only depend on the angle $\vartheta$ between the photon vectors ${\bf k}$ and ${\bf k}_{1}$  then we could say that there is the angle symmetry of the two-photon correlation spectra. This symmetry, as far as we know, is experimentally observed for the positron annihilation in solid targets. However, this appears to be due to the fact that, the positron energy and momentum relaxations are more rapid processes compared with the annihilation process in solids. 
\par We draw attention to two effects. The first is that the observables \eqref{68} and \eqref{75} depend on the momentum of free particles, and, hence, on the initial energy of the electrons and positrons. Taking into account \eqref{72}, we can argue that the ${\bf p}_{\bot}$ dependence of the observables will result in the first contribution to the asymmetry of the two-photon angular correlation spectra.
\par The second effect is due to the radiation pattern of the massless bosons because, according to \eqref{66}, the cross section is proportional to $\cos ^{2} \theta_{g}$. That is, these bosons are emitted predominantly along the colliding axis of two bunches. This results in the second contribution to the asymmetry of these spectra.  
\subsection{Procedure of numerical calculations}
Due to computational constraints, we were not able to investigate these two effects presented above. The aim of the subsequent part of this work is to find out the characteristic form of the two-photon angular correlation spectra which are due to the presence of the third particle that is the massless boson. We are trying to answer the question: what typical widths of these spectra can be expected for the reaction \eqref{57}?
\par With this aim, we replace $\cos^{2} \theta_{g} \to 1/2$. In the low-energy limit of the initial particles we can use 
$k_{\bot}>>p_{\bot}$. In this case the observables \eqref{68} and \eqref{75} depend only on the angle $\vartheta$ between the wave vectors of the two photons emitted. As a result, the cross section (68) is reduced to the form:
\begin{equation} \label{76} 
\frac{\partial^{2}\sigma}{\partial \Omega_{{\bf k}}\partial \Omega_{{\bf k}_{1}}} =
(4\pi)^{-3}r_{e}^{2} \frac{v_{e}}{c} G(\cos \vartheta) 
\end{equation} 
\noindent Here we use notations:
\begin{equation} \label{77} 
G(\cos \vartheta)=\int _{0}^{k_{e}} F(k,\cos \vartheta) \left|\chi(k_{\bot}(k,\cos \vartheta);g(k,\cos \vartheta))\right|^{2}dk, 
\end{equation} 
where
\begin{equation} \label{78} 
F=\frac{k_{e} k(k_{e}-k)}{(2k_{e}-k(1-\cos \vartheta))^{2}},
\end{equation} 
\noindent In \eqref{77} and \eqref{78}, $g$ and $k_{\bot}$ are given by \eqref{69} and \eqref{71}, respectively.
\par Taking into account \eqref{75}, $\zeta$ is rewritten as:
\begin{equation} \label{79} 
\zeta(\cos \vartheta)=(4\pi)^{-3} \frac{r_{e}^{2}}{c} G(\cos \vartheta) 
\end{equation} 
\par Further it is convenient to introduce dimensionless variables. For non-relativistic electrons and positrons  
$k_{e}={\mathchar'26\mkern-10mu\lambda}_{e}^{-1}$. Hence, the dimensionless wave function of the transverse motion of the coupled pair is ${\mathchar'26\mkern-10mu\lambda}_{e}^{-1}* \chi \to \chi$, the photon and boson energies become  
${\mathchar'26\mkern-10mu\lambda}_{e} k \to k$ and ${\mathchar'26\mkern-10mu\lambda}_{e}g\to g$.
As a result, the functions (69), (71) and (78) are dimensionless:
\begin{equation} \label{80} 
g=\frac{2 -k(2 -k)(1-\cos \vartheta )}{2-k(1-\cos \vartheta)}
\end{equation} 
\begin{equation} \label{81} 
k_{\bot} =\frac{2k(1-k)\sqrt{1-\cos^{2} \vartheta}}{2 -k(2 -k)(1-\cos \vartheta)},
\end{equation} 
\noindent and
\begin{equation} \label{82} 
F=\frac{k(1 -k)}{(2 -k(1-\cos \vartheta ))^{2}}.
\end{equation} 
\par The quantities $\sigma$ and $\zeta$ are also given by \eqref{76} and \eqref{79}, respectively, 
and the function \eqref{77} takes the form:
\begin{equation} \label{83} 
G(\cos \vartheta)=\int_{0}^{1}F(k,\cos \vartheta) \left|\chi(k_{\bot}(k,\cos \vartheta);g(k,\cos \vartheta))\right|^{2}dk. 
\end{equation} 
 \par Both the observables \eqref{76} and \eqref{79}  are defined by the same function \eqref{77}. The latter is determined by the 
momentum-space wave function of the transverse motion of the coupled electron-positron pair, $\chi(k_{\bot};g)$.
As was found above, for small momenta of the boson, $g<<m$, this wave function satisfies the homogeneous Fredholm integral equation of the second kind \eqref{31} with the non-Fredholm kernel  \eqref{32}-\eqref{34}. We were able to calculate the transverse wave function only in this case. It is a fortunate coincidence that the small energies of the massless bosons, $g<<m$, are of fundamental importance for the reaction \eqref{57}, as will be demonstrated below. 
\par Firstly, to find the wave functions of the massless composite boson, $\chi(f;g)$, we used exactly the calculation procedure presented in Section 2.3 with the same $N=4501$ and $f_{0}(g)=100*(1+g/\alpha)$. Then, using the obtained wave function 
$\chi (f;g)$, for the given $k$ and $\vartheta$  the value of this wave function  
$\chi (k_{\bot}(k,\cos \vartheta);g(k,\cos \vartheta))$ was calculated by formula:  
\begin{equation} \label{84} 
\chi(k_{\bot};g)=\frac{\alpha}{g} \int_{0}^{f_{0}(g)} Q(k_{\bot},f) \chi (f;g)df,
\end{equation} 
\par Finally, to find the observables \eqref{76} and \eqref{79} for the given angle $\vartheta$, the integration over 
$k$ in \eqref{83} was carried out. The angular dependence of  $\zeta (\cos \vartheta)$ gives us the $2\gamma$ angular correlation spectra for the reaction \eqref{57}. 
\subsection{Results}
The dependence of the boson energy $g$ on both the photon energy $k$ and angle $\vartheta$ is determined by \eqref{80}. 
The given angle can be realized for different boson energies, since the sums $k+k_{1}+g=2$ and ${\bf k}+{\bf k}_{1}+{\bf g}=0$ 
are only fixed. However, for each angle there is a minimum value of the boson kinetic energy. This minimal energy which can easily be found from \eqref{80}, is 
\begin{displaymath}
g_{min}=2tg(\phi/2)\left[ (1+tg^{2}(\phi/2))^{1/2}-tg(\phi/2)\right]
\end{displaymath}  
Here we introduce the angle $\phi=\pi-\vartheta$ that will be used below. 
\par Fig. 7 presents the dependencies of the boson energy $g$ converted to keV, on the photon energy $k$ for three angles $\phi$. 
The curve 1 corresponds $\phi=$0.146 mrad, for which $g_{min}=$74.6eV. For the angle $\phi$=0.431 mrad this minimal energy is equal to $g_{min}=$220eV, as shown by the curve 2 on Fig. 7, and for the last curve 3 $g_{min}=$511eV at $\phi$=1 mrad.  
As follows from Fig. 7, each $g_{min}$ is corresponded to the photon energy $k_{*}$. For the given angle $\phi$ and the boson energy $g>g_{min}$ there is a pair of the photon energy values, as is clearly demonstrated by the horizontal line on Fig. 7. 
One value of the photon energy from this pair is greater than $k_{*}$, and the other is less than $k_{*}$.      
These paired values of photon energies will contribute to the integral on the right-hand side of \eqref{83}.
We show below that this integral is evaluated in some neighborhood of $k_{*}$.   
\begin{figure}[ht]
\centering
\includegraphics[width=8.5cm]{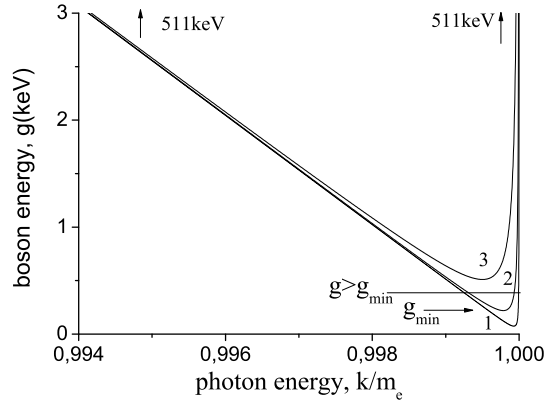}
\caption{
The photon energy dependence of the boson energy at given angle between the momenta of two photons emitted. 
For the curves from 1 to 3: 1 - $\phi$=0.146 mrad; 2 - $\phi$=0.431 mrad; 3 - $\phi$=1 mrad. 
\label{f7}}
\end{figure}
\par There are two more functions, namely $k_{\bot}$, \eqref{81}, and $F$, \eqref{82}, which are also important for calculating the observables \eqref{76} and \eqref{79}. The photon energy dependencies of these quantities are presented in Fig. 8 
for the same angles as in Fig. 7. The value of $k_{\bot}$ goes to zero in the limit $k\to 0$, reaches a maximum, and then vanishes in the limit $k\to 1$. In fact, Fig. 7 is shown the three dependencies of $F(k)$ for the three angles. At small angles they are very close to each other. They differ in that for different values of the photon energy $\frac{2}{3-cos(\phi)}$ 
the curves reach the maximum values $\frac{1}{8(1-cos(\phi)}$, and then turn sharply to zero in the limit $k\to 1$.
\begin{figure}[ht]
\centering
\includegraphics[width=8.5cm]{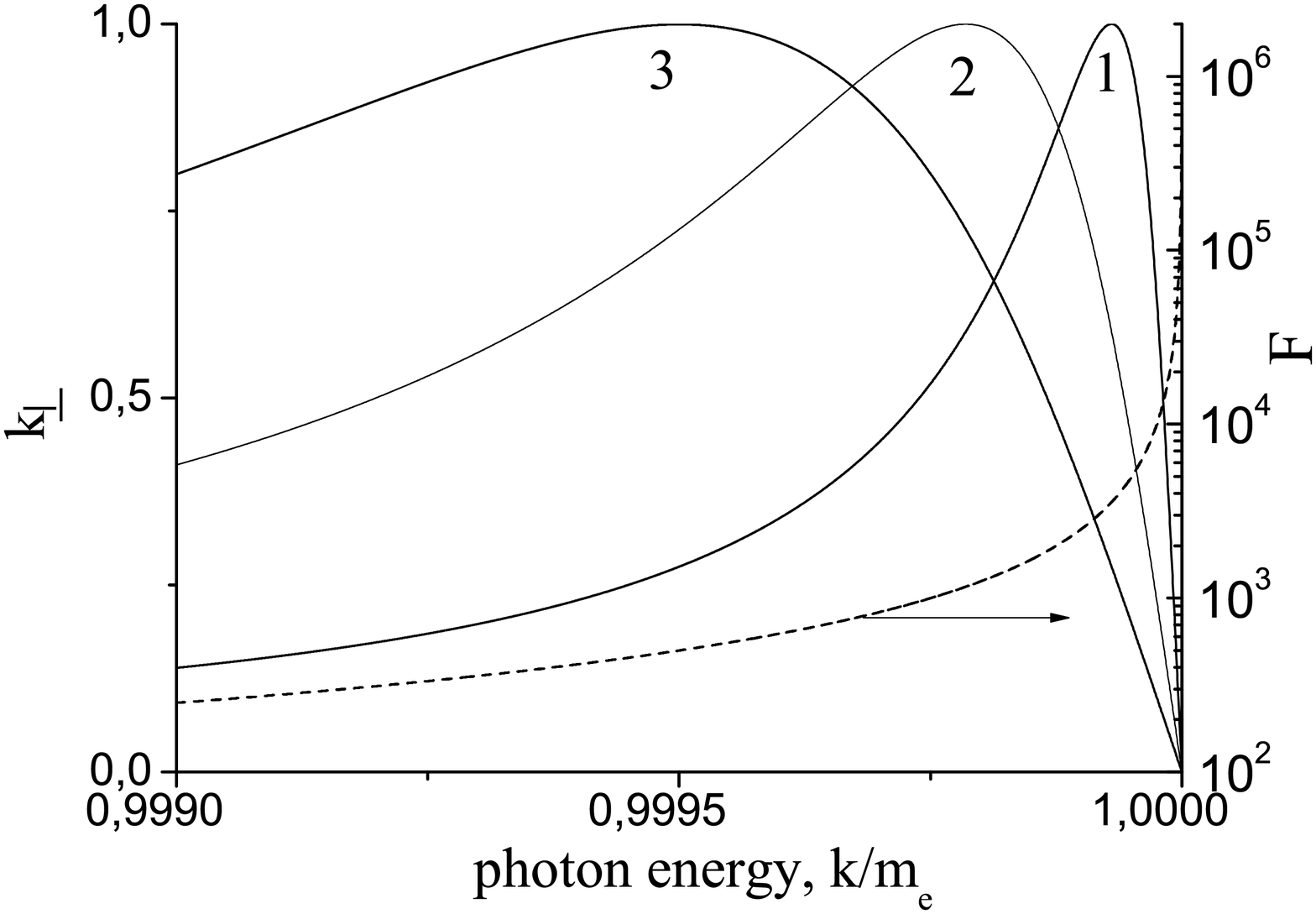}
\caption{
The photon energy dependence of the functions $k_{\bot}$ and $F$. The angles $\phi$ are the same as in Fig. 6.   
\label{f8}}
\end{figure}
\par The angular dependence of the observables is determined by the same integral \eqref{83}. The $k$ dependencies of the integrand function, $F(k,-\cos \phi) \left|\chi(k_{\bot}(k,-\cos \phi);g(k,-\cos \phi))\right|^{2}$, are demonstrated in Fig. 9-12for four angles $\phi$.
\par We begin with the angles $\phi$ which give a very small contribution to the $2\gamma$ angular correlation spectrum for the reaction \eqref{57}. Fig. 9 presents the integrand as a function of the photon energy $k$ at the angle $\phi=$1 mrad. As shown above, the transverse wave function $\chi(f,g)$ is characterized by anharmonic oscillations in the $f-$ momentum space. As a result, the functions $F\left|\chi\right|^{2}$ always represent irregular changes. Except for one point corresponded to 
$g_{min}$, the all other points on the curve  presented in Fig. 3, are paired because the horizontal line $g=const$ intersects the curve $g(k)$ in the two points, as shown in Fig. 7. 
\begin{figure}[ht]
\centering
\includegraphics[width=8.5cm]{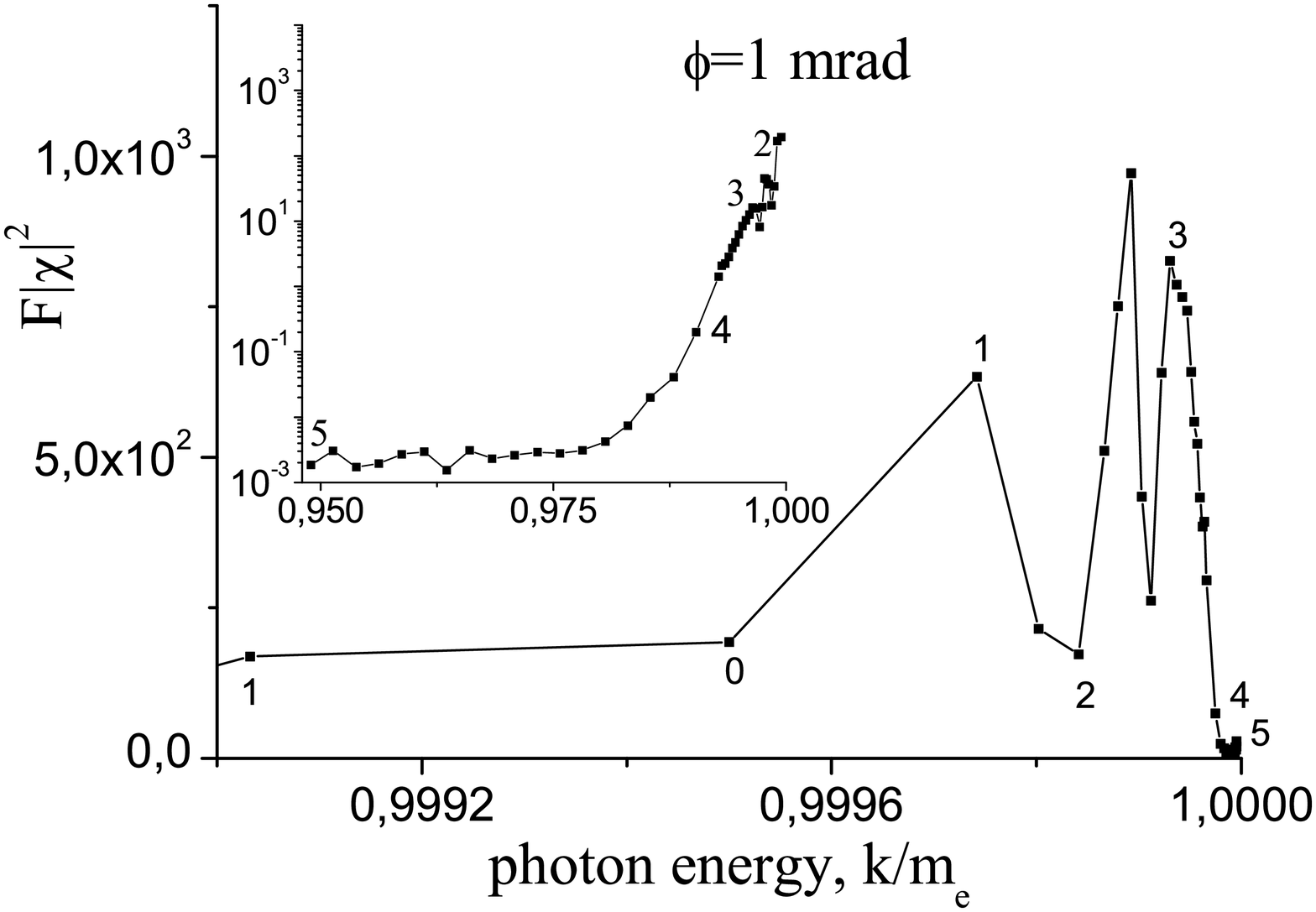}
\caption{
The integrand $F\left|\chi\right|^{2}$ as a function of the photon energy $k$ for the angle $\phi=$1 mrad and 
$k\geq k_{0}=0.999$. Inset: The same except for $k\leq k_{0}$. For the labeled points the energies of massless bosons emitted are given in the text.   
\label{f9}}
\end{figure}
\par The massless boson energies that correspond to the marked points in Fig. 9, are as follows:
for the point 0 the energy of emitted boson equal to $g_{min}=$511eV; for the paired points 1 the boson energy $g =$626eV, 
2 - 884eV, 3 - 1.86keV, 4 - 4.98 keV, and the last pair points 5 - 26.10keV. Thus, the energies of the emitted massless bosons are much smaller than the electron mass. 
\par For the right-hand point 5, the dimensionless photon energy differs from 1 by $5\times10^{-6}$. Therefore, for the larger photon energies contribution to the integral in the right-hand sides of \eqref{83}  is very small. Using the data presented in Fig. 9, we have found that $G=$0.40 at $\phi=$1 mrad.
\par Fig. 10 shows the integrand $F\left|\chi\right|^{2}$ as a function of the photon energy  $k$ at the angle $\phi=$0.252 mrad. The characteristic values of this function have significantly increased in comparison with that presented in Fig. 9 for the angle 
$\phi=$1 mrad. 
\begin{figure}[ht]
\centering
\includegraphics[width=8.5cm]{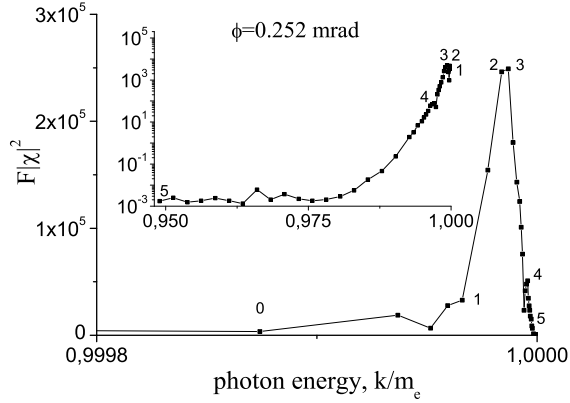}
\caption{
The same as in Fig. 8 except for the the angle $\phi=$0.252 mrad and $k_{0}=0.9998$.
\label{f10}}
\end{figure}
\par For the angle $\phi=$0.252 mrad, the energies of the emitted composite bosons for the marked points in Fig. 10 are:   
the point 0 corresponds to the energy $g_{min}=$129eV, the paired points 1 correspond to $g =$256eV, 2 - 511eV, 3 - 626keV, 
4 - 1.86keV, and the last pair points 5 - 26.10keV. 
\par For the right-hand point 5, the dimensionless photon energy differs from 1 by $3\times10^{-7}$. Therefore, contribution to the integral in the right-hand sides of \eqref{40} is very small for the larger photon energies. We have calculated that $G=$11.07 at $\phi=$0.252 mrad.
\par Note that due to computational constraints, we could not find the eigenfunctions of the kernel \eqref{33} - \eqref{34} for the boson energies significantly less 130 eV. For this reason, in subsequent results calculated for smaller angles $\phi$,  the contribution to the $G$ function due to the emission of massless bosons with energies less than 129 eV is not taken into account. Obviously, this will lead to some underestimation of the $G$ function. 
\par The photon energy dependence of the integrand function in the right-hand side of \eqref{83} at the angle $\phi=$0.12 mrad is shown in Fig. 11. For this angle the point 0 would correspond to the boson energy $g_{min}=$61.3eV. However, we could not calculate the massless boson wave function for such low energy. Therefore, this point is absent in Fig. 11.
The paired points 1 correspond to $g =$129eV, 2 - 220eV, 3 - 373keV, 4 - 1.02keV, and the last point 5 - 26.10keV (the left point 5 at the photon energy $k=0.949$ is not shown). The two arrows shown in Fig. 11 and the inset, represent the photon energy region in which the integrand in Eq. \eqref{40} was not determined and, accordingly, integration  over this region 
$0.999764<k<0.999985$ was not carried out. The value of $G$ was found to be equal to $G=$18.66 at the angle $\phi=$0.12 mrad. 
\begin{figure}[ht]
\centering
\includegraphics[width=8.5cm]{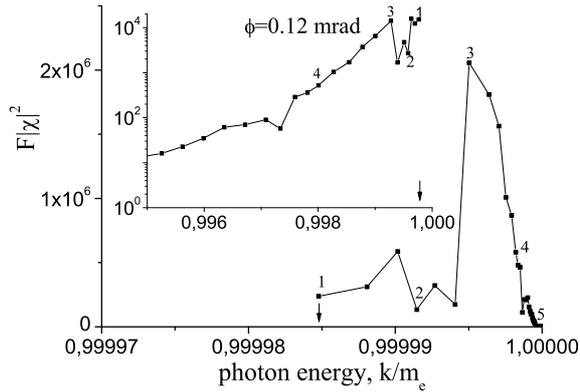}
\caption{
The same as in Fig. 8 except for the the angle $\phi=$0.12 mrad and $k_{0}=0.99997$.
\label{f11}}
\end{figure}
\par With account for the definitions \eqref{80}-\eqref{82}, the singular point of the integral \eqref{83} is the angle 
$\vartheta=\pi$. It means that $\phi=0$, and in the reaction \eqref{57}, along with the boson the two photons having  different energies, are emitted at 180$^{0}$ to each other. In this case  $g=1-k$, $k_{\bot}=0$ and $F=\frac{k}{4(1-k)}$. Then the integral \eqref{83} takes the form:
\begin{equation} \label{85} 
G(-1)=\int_{0}^{1} \frac{kdk}{4(1-k)} \left|\chi(0;1-k)\right|^{2}, 
\end{equation} 
where the function $\chi (0;1-k)$ is found from the equation:
\begin{displaymath} 
\chi (0;1-k)=\frac{2\alpha}{1-k}\int_{0}^{\infty} \left(\frac{\Theta(f-1)}{\sqrt{f^2-1}}+
i\frac{\Theta(1-f)}{\sqrt{1-f^2}}\right)f\chi (f;1-k)df,
\end{displaymath} 
\noindent where $\Theta$ is the Heaviside step function.
\par Note that in the limit $k\to 1$ the logarithmic divergence in \eqref{85} is absent, since the function $\chi(f,0)$ vanishes 
for any $f$. It is due to the fact that the massless boson can not be in the state of rest ($g\neq 0$).
\par The integrand in Eq. \eqref{85} is presented in Fig. 12 for the photon energy $k\geq 0.99$. We do not show this function
in the region $0.948 \leq k \leq 0.99$ where it gives very small contribution to $G(-1)$. 
The point labeled 1 corresponds to the energy of the massless boson $g=129$eV. In the remaining marked points the boson energies are the following: for point 2 - 220eV, 3 - 626eV, 4 - 1.86keV and 5 - 2.98keV.
\begin{figure}[ht]
\centering
\includegraphics[width=8.5cm]{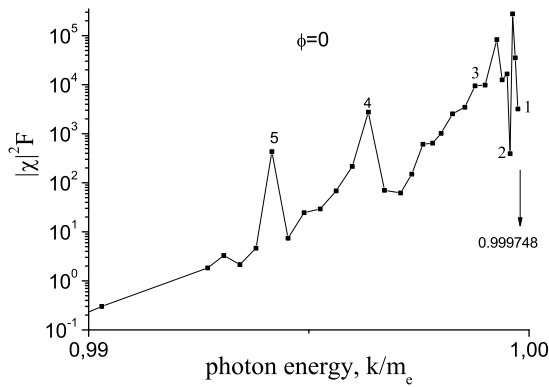}
\caption{
The integrand function in \eqref {85} for the angle $\phi=0$ in the region of the photon energies 
$0.99\leq k\leq 0.999748$.
\label{f12}}
\end{figure}
\par The vertical arrow in Fig. 12 shows the limitation of the region in which the integration in \eqref{85} is carried out,
so that the region $0.999748<k<1$ remains unaccounted for. The presented curve in Fig. 12 leads to the value $G(-1)$=45.72. 
Of course, the value is underrated because events in which the products of the reaction \eqref{57} are the massless bosons with energies less than $129$eV, were not taken into account.
\par Summarizing all the results obtained in this study, we have found the $2\gamma$ angular correlation spectrum shown 
in Fig. 13. Kinks on the line at $\phi \simeq \pm 0.22$mrad indicate the boundary of the small-angle region with the understated data for this spectrum. The reason for this understatement was discussed above. It can be seen that for the reaction \eqref{57} the $2\gamma$ angular correlation spectrum is characterized by a narrow peak with the full-width-at-half-maximum (FWHM) not exceeding 0.2 mrad. 
\begin{figure}[ht]
\centering
\includegraphics[width=8.5cm]{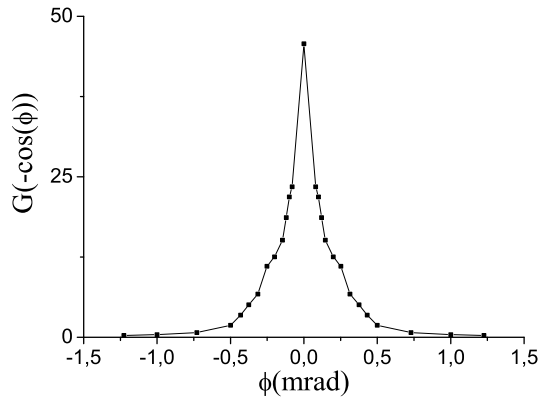}
\caption{
The $2\gamma$ angular correlation spectrum for the reaction \eqref{57}.  
\label{f13}}
\end{figure}
\section{Conclusion}
In the present work we treated the electron and positron as independent particles, each being characterized by the complete set of the Dirac plane waves. Although this treatment is beyond the standard QED theory, it does not contradict the free particle Dirac theory because, at least, the electron and positron are fermions.
\par This approach leads inevitably to another choice of the free particle propagator with compared to that being currently used in QED. Then the bound-state Bethe-Salpeter equation in the ladder approximation with these free propagators was investigated. 
The branch of the massless composite bosons formed by the bound electron-positron pairs with the actual coupling equal to the fine structure constant states, have been found. Summarizing the results presented in Fig. 1-6, for the case of the \textit{S}-state of the coupled electron-positron pair and small momenta of the composite boson, $g<<m$, the following conclusions can be made: 1) The massless boson wave functions are complex-valued and normalized; 2) The average distance between the electron and positron diverges as the boson kinetic energy goes to zero that explains that the massless compound boson cannot be at rest;
3) The momentum wave function extension leading to the increase of the average transverse momentum of the coupled pair, and the coordinate wave function compression accompanying with the decrease of the average transverse distance between the electron and positron, are continuously occurred with increasing the boson energy.  
\par In accordance with the central particle-antiparticle concept in QED, products of the low-energy electron-positron annihilation  are just a few photons, and nothing remains of the electron and positron. The singlet pair of the free particles with the center of mass at rest is, with the greatest probability, converted into two photons, which, due to the momentum conservation, should be emitted at the angle 180$^0$ to each other. The treatment of the electron and positron as independent particles leads to the reaction \eqref{57} in which together with emitted photons the reaction products involve the massless boson formed by the coupled electron and positron pair. For this reaction the $2\gamma$ correlation spectra must always have the finite angular width since the three particles are emitted. That is, the presence of the width is fundamental. As was found above, for the extremely non-relativistic particles the two-photon angular correlation spectrum is characterized by a narrow line with the full-width-at-half-maximum (FWHM) not exceeding 0.2 mrad. 
\par In many known experiments on the low-energy positron annihilation in condensed and gaseous matters (see 
\cite{Pus,Col,Gri,Ste,Cha} and references therein) the center of mass of annihilating pairs is usually in motion with respect to an observer. Then the angle between the photon directions departs from 180$^{0}$ by an amount of the order of 
$v_{cm}/c$, where $v_{cm}$ is the velocity of the center of mass and $c$ is the velocity of light, and, hence, the annihilation angular correlation spectra of the $2\gamma$ radiation are characterized by finite widths as well. 
Obviously, in these experiments it is not possible to distinguish the reaction \eqref{57} from the conventional annihilation. In the initial state (left-hand side of \eqref{57}) the electron and positron can occur in a parapositronium state that defines other channel of this process.    
\par Note that the determination of which of the two discussed reactions takes place actually, is possible. For this purpose,
experiments with colliding beams of extremely non-relativistic electrons and positrons should be carried out. To detect the reaction \eqref{57}, parameters of the colliding beams are very important. The ideal would be mono-energetic colliding beams of the annihilating particles with the center of mass at rest. However, the spectra always have finite widths, $\delta \varepsilon$, which can be considered as the same both the electron and positron beams. Then, it is easy to estimate that for the conventional annihilation in vacuum the typical width of the $2\gamma$ angular spectrum is of the order of 
\begin{displaymath}
\phi_{r}\simeq \frac{v_{e}}{c}\frac{\delta \varepsilon}{\varepsilon},
\end{displaymath}
where $v_{e}$ and $\varepsilon$ is the average velocity and energy of the particles. 
\par For the reaction \eqref{57} the angular width of the two-photon correlation spectrum is of the order of $10^{-4}$rad. Then one can estimate the required parameters of the beams:
\begin{displaymath}
\frac{v_{e}}{c}\frac{\delta \varepsilon}{\varepsilon}<10^{-4}.
\end{displaymath}
\noindent The angular resolution of two-photon detectors should be less then $\phi_{r}$.
\par For the electron and positron beams with the average energies $\varepsilon=3$eV and the spectral line widths 
$\delta \varepsilon=50$meV we obtain $\phi_{r}=0.6\times 10^{-4}$ that satisfies the condition required. 
The production of the positron beam with close parameters $\varepsilon=1$eV and $\delta \varepsilon=75$meV (for which
the value of $\phi_{r}$ is slightly larger, $\phi_{r}=1.5\times 10^{-4}$) was reported in Ref \cite{Lyn}.
\par At high energies, say even for keV beams (not to mention MeV beams), it is difficult to obtain such narrow lines with 
$\frac{\delta \varepsilon}{\varepsilon} \simeq 10^{-4}$. 
\par Note that in the literature we did not find similar experiments with such colliding low-energy beams in vacuum. We emphasize that it is very important to use exclusively non-relativistic particle beams with energies of a few eV, as discussed below. Exactly for such energies of the particles one can obtain the necessary energy homogeneity of the colliding beams. 
\par Unlike the conventional annihilation, for the reaction \eqref{57} the two-photon correlation spectrum has a finite angular width in principle. This can be used to establish the actual process. Consider such colliding beams of the low-energy spin-polarized electrons and positrons, and assume that the angular resolution of two-photon coincidence-count detector is less than $\phi_{r}$. In the case of observation of a narrow peak the angular width of which is determined by the instrument resolution, one can conclude that the conventional annihilation takes place only, the positron is the antiparticle of the electron, and, respectively, our approach that the electron and positron are independent particles, is not correct. If the angular width of the 
$2\gamma$ correlation spectrum  will be greater than the instrument resolution that should be $<10^{-4}$rad, then it will mean that the composite massless bosons formed by the coupled electron-positron pairs, do really exist, and the reaction \eqref{57} does occur. 

\end{document}